\documentclass[11pt]{article}

\usepackage{amsfonts}
\usepackage{amsmath}
\usepackage{amsthm}
\usepackage{amssymb}
\usepackage{dsfont}
\usepackage{apacite}
\usepackage{epsfig}
\usepackage{natbib}
\usepackage{setspace}
\usepackage{verbatim}

\theoremstyle{plain}
\newtheorem{theorem}{Theorem}
\newcommand{\ind}{{\mathds{1}}}
\DeclareMathOperator{\Prob}{\mathbb{P}}
\DeclareMathOperator{\Exp}{\mathbb{E}}
\DeclareMathOperator{\Var}{\mathbb{V}}
\newcommand{\bs}{\boldsymbol}
\DeclareMathOperator{\logit}{logit}

\newcommand{\cv}{\text{\small\sc cv}}
\newcommand{\tr}{{\text{\sc t}}}

\begin{document}

\title{Importance sampling for weighted binary random matrices with specified margins}

\author{Matthew T. Harrison$^*$ \\ Jeffrey W.~Miller\footnote{Matthew T.~Harrison is Assistant Professor of Applied Mathematics, Brown University, Providence, RI 02912 (email: Matthew\_Harrison@Brown.edu) and Jeffrey W.~Miller is a graduate student of Applied Mathematics, Brown University, Providence, RI 02912 (email: Jeffrey\_Miller@Brown.edu). This work was supported in part by the National Science Foundation (NSF) grant DMS-1007593, the Defense Advanced Research Projects Agency (DARPA) contract FA8650-11-1-7151, and, while MTH was in the Department of Statistics at Carnegie Mellon University, by the NSF grant DMS-0240019 and the National Institutes of Health (NIH) grant NIMH-2RO1MH064537.  Any opinions, findings, and conclusions or recommendations expressed in this material are those of the authors and do not necessarily reflect the views of the NSF, the NIH,  or DARPA.  The authors thank Sam Kou for sharing his code for approximating $\alpha$-permanents.} \\ Division of Applied Mathematics \\ Brown University \\ Providence, RI 02912} 

\maketitle



\begin{abstract} A sequential importance sampling algorithm is developed for the distribution that results when a matrix of independent, but not identically distributed, Bernoulli random variables is conditioned on a given sequence of row and column sums.  This conditional distribution arises in a variety of applications and includes as a special case the uniform distribution over zero-one tables with specified margins.  The algorithm uses dynamic programming to combine hard margin constraints, combinatorial approximations, and additional non-uniform weighting in a principled way to give state-of-the-art results.  
\end{abstract}


\section{Introduction} \label{s:intro}

Let $\Omega^*$ denote the set of $m\times n$ binary matrices with row sums $\bs r=(r_1,\dotsc,r_m)$ and column sums $\bs c=(c_1,\dotsc,c_n)$, and let $\bs w=(w_{ij})\in[0,\infty)^{m\times n}$ be a given nonnegative matrix.  Define the distribution $P^*$ on $\{0,1\}^{m\times n}$ via
\begin{equation} P^*(\bs z) = \frac{1}{\kappa} \prod_{ij} w_{ij}^{z_{ij}} \ind\{\bs z\in \Omega^*\} , \quad \quad \kappa = \sum_{\bs z\in\Omega^*} \prod_{ij} w_{ij}^{z_{ij}} , \label{e} \end{equation}
where $\ind$ is the indicator function and where we assume $\kappa > 0$.  $P^*$ is the conditional distribution of an $m\times n$ array of independent Bernoulli random variables, say $\bs B=(B_{ij})$, with $\Prob(B_{ij}=1)=w_{ij}/(1+w_{ij})$ given the margins $\bs r$ and $\bs c$, where $\Prob$ denotes probability.
This paper describes an importance sampling algorithm that can be used for Monte Carlo approximation of probabilities and expectations under $P^*$ and also for Monte Carlo approximation of $\kappa$.  After a preprocessing step, sampling from our proposal distribution requires $O(md)$ operations per matrix, where $d=\sum_j c_j=\sum_i r_i$ is the total number of ones in the matrix.  

We are not aware of any existing importance sampling algorithms that permit practical inference under $P^*$, although many special cases have been studied in the literature.  For example, if $\bs w\equiv 1$ then $P^*$ is the uniform distribution over zero-one tables with specified margins, or equivalently, the uniform distribution over bipartite graphs with specified degree sequence.  For square matrices, if $\bs w$ is identically one except with a zero diagonal, then $P^*$ corresponds to the uniform distribution over directed graphs with specified degrees.  And if $\bs r = \bs c\equiv 1$, then $P^*$ is a distribution over weighted permutation matrices and $\kappa$ is the permanent of $\bs w$.  Empirically, our algorithm outperforms all existing importance sampling algorithms in these special cases.  Although our algorithm works well for most examples arising in practice, performance depends on $\bs r$, $\bs c$, and $\bs w$.  Highly irregular margins or highly variable $\bs w$, particularly many zero entries in $\bs w$, tend to cause poor performance.

$P^*$ factors in such a way that we need only focus on the distribution, say $P$, of the first column.  The columns are sampled sequentially, with each successive column viewed as the first column of a smaller matrix with updated margins based on the previously sampled columns.  We decompose the structure of $P$ into margin constraints, combinatorial factors, and non-uniform weighting terms, combine approximations of these terms in a principled way, and then use a dynamic programming algorithm to exactly and efficiently sample from the resulting proposal distribution $Q$ for the first column.  Sequentially sampling columns in this way defines a proposal distribution $Q^*$ for the whole matrix.  This strategy for algorithm design works well for many similar problems, including symmetric matrices and nonnegative integer-valued matrices, each of which will be described elsewhere owing to space constraints.  It seems likely that the design principles used for our approach are applicable much more broadly.

\section{Motivating applications}

\subsection{Conditional inference for graphs and tables} \label{s:conditional}

Let $\bs B\in\{0,1\}^{m\times n}$ be a matrix of independent Bernoulli random variables with
\begin{equation} \textstyle \logit \Prob(B_{ij}=1) = \alpha_i + \beta_j + \sum_k \theta_k \xi_{kij} , \label{e:logit} \end{equation}
where $\bs\alpha$, $\bs\beta$, and $\bs\theta$ are parameters, perhaps with constraints to ensure identifiability, and $\bs \xi$ is a collection of observed covariates.  Models of this form arise, for example, in educational testing, where $B_{ij}$ indicates whether or not subject $i$ responded correctly to question $j$.  If $\bs\theta\equiv 0$, then the model reduces to the classical Rasch model \citep{rasch1960probabilistic,rasch1961general}.  Otherwise, it is an extension of the Rasch model to include item-specific covariate effects, such as each subject's prior exposure to the content being tested in each question.  This model is also a simple version of models used for the analysis of network data \citep[c.f.,][]{holland1981exponential,fienberg1985statistical,goldenberg2010survey} where $\bs B$ is the adjacency matrix of a directed graph, $\bs \alpha$ and $\bs\beta$ allow for degree heterogeneity, and $\bs \xi$ is a collection of edge-specific covariates.  For example, for social network data we might have that $B_{ij}$ indicates whether subject $i$ reported subject $j$ as a friend, $\alpha_i$ controls the relative propensity for subject $i$ to report friends, $\beta_j$ controls the relative propensity for subject $j$ to be reported as a friend, and $\xi_{kij}$ indicates whether the relationship between subject $i$ and $j$ is of type $k$.

In both of these examples, if $\bs\theta$ is the only parameter of interest, then the nuisance parameters $\bs\alpha$ and $\bs\beta$ complicate inference and can be removed by conditioning on the row and column sums of $\bs B$ \citep[e.g.,][]{cox1958regression,holland1981exponential,mehta1995exact,harrison2012conservative}.  Conditioning also results in inferential procedures that are robust to modeling assumptions, implicit in \eqref{e:logit}, about the distribution of the margins.  The resulting conditional model that drives inference is exactly $P^*=P^*_\theta$ with $w_{ij}=\exp(\sum_k \theta_k \xi_{kij})$, perhaps with the additional constraint that $w_{ii}=0$ in the case of network data.  The conditional model is a natural exponential family in $\bs \theta$ with no nuisance parameters, but with an intractable normalization constant $\kappa=\kappa_\theta$.  \citet[Example 4.2]{harrison2012conservative} provides details about an importance sampling approach to exact conditional inference for this model.  The example there is based on a preliminary version of the algorithm presented here.  

Other approaches to conditional inference in this setting include exhaustive enumeration, such as the algorithms for conditional logistic regression in Stata \citep{stata} and LogXact \citep{logxact}, Markov chain Monte Carlo approaches, such as the {\tt elrm} R package \citep{zamar2007elrm}, and analytic approximations, such as the {\tt cond} R package \citep{cond,brazzale2008accurate}, none of which are practical for larger matrices and/or multivariate $\bs\theta$.  Approximation of $\kappa$ was considered in \citet{barvinok2010number}.  

\subsection{The uniform distribution and model validation}

If $\bs w\equiv 1$, or more generally, if $w_{ij}=\exp(\alpha_i+\beta_j)$ for real-valued $\bs \alpha$ and $\bs \beta$, then $P^*$ is the uniform distribution on $\Omega^*$.  The uniform distribution can be used for testing if $\bs\theta\equiv 0$ in model \eqref{e:logit}, or equivalently, for model validation of \eqref{e:logit} specialized to the case of $\bs\theta\equiv 0$.  Prominent examples include goodness-of-fit tests for the Rasch model \citep[e.g.,][]{rasch1960probabilistic,rasch1961general,ponocny2001nonparametric,chen2005exact,Chen:Sequential:2005} and for random bipartite graphs and directed graphs without reciprocity \citep[e.g.,][]{wasserman1977random,holland1981exponential,snijders1991enumeration}.  

The uniform distribution over $\Omega^*$ also plays a central role in testing for the presence of interactions in co-occurrence tables, particularly co-occurrence tables arising in ecology, where $B_{ij}$ indicates the existence of species $i$ in location $j$ \citep[e.g.,][]{connor1979assembly,snijders1991enumeration,gotelli2000null,Chen:Sequential:2005}.  In these contexts, the uniform distribution subject to the margin totals is taken as a null hypothesis of no interaction among species (without necessarily assuming model \eqref{e:logit}).  Our original motivation for developing these algorithms came from a similar problem in neuroscience where $B_{ij}$ indicated whether neuron $i$ produced an action potential in time bin $j$, and the uniform distribution was taken as a null hypothesis of a lack of interaction among the neurons.  This can be viewed as an example of conditional testing for multivariate binary time series.  In that example, $mn\approx 10^8$ and existing algorithms in the literature were not practical.    

Besides the many statistical applications, when $\bs w\equiv 1$, the normalization constant $\kappa$ is the number of binary matrices with specified margins, a topic of enduring interest in theoretical computer science \citep[e.g.,][]{kannan1999simple,jerrum2004polynomial,bezakova2007sbc} and combinatorial approximation \citep[e.g.,][]{bekessy1972asymptotic,mckay1984amp,greenhill2006aes,canfield2008aed,barvinok2010matrices}.  Importance sampling algorithms for $P^*$ can be used to provide efficient approximations of $\kappa$ \citep{blanchet2006isa}.

Monte Carlo sampling algorithms for the uniform distribution have been developed by many authors \citep[e.g.,][]{besag1989gmc, mckay1990uniform,snijders1991enumeration,rao1996markov,Chen:Sequential:2005, blanchet2006isa, bezakova2007sbc, chen2007cit, verhelst2008ema, Bayati:Sequential:2009}.  The approach here was inspired by the importance sampling algorithm in \citet{Chen:Sequential:2005}, but provides a more principled method for algorithm design that leads to substantial improvements in the uniform case and that also extends to the non-uniform case. 

\subsection{Permanents and permanental processes}

If $\bs w$ is square and $\bs r=\bs c\equiv 1$, that is, if $\Omega^*$ is the set of permutation matrices, then $\kappa$ is the permanent of $\bs w$, also of enduring interest in theoretical computer science \citep[e.g.,][]{valiant1979complexity,jerrum2004polynomial}.  A variety of generalizations of permanents and determinants can be expressed as $\kappa\mu$, where $\mu=\Exp(h(\bs Z))$ for some function $h$, where $\Exp$ denotes expected value, and where $\bs Z$ has distribution $P^*$ \citep[e.g.,][]{littlewood1950theory,vere1988generalization,vere1997alpha,diaconis2000immanants}.  In principle, the algorithms here could be used to approximate the value of any of these objects, but the practicality of this approach depends heavily on $h$.  For example, the $\alpha$-permanent \citep{vere1988generalization,vere1997alpha} is  
\begin{equation} \text{per}_\alpha(\bs w) = \kappa \Exp(\alpha^{\text{cyc}(\bs Z)}) , \label{e:perm} \end{equation}
where $\alpha\in\mathbb{R}$ and $\text{cyc}(\bs z)$ is the number of disjoint cycles in the permutation corresponding to $\bs z$.  The case $\alpha=1$ corresponds to the permanent of $\bs w$, and the case $\alpha=-1$ corresponds to $(-1)^n$ times the determinant of $\bs w$.  Permanents and $\alpha$-permanents arise in probability, statistics, and statistical physics in connection to permanental processes and random fields \citep[e.g.,][]{macchi1975coincidence,diaconis2000immanants,shirai2003random,mccullagh2006permanental,kou2009approximating} and the distribution of order statistics \citep{vaughan1972permanent,bapat1989order}.  Our approach is often effective for approximating \eqref{e:perm} when $\alpha>0$ and $|\log\alpha|$ is small.

\section{Algorithm design}

\subsection{The target distribution for the first column} \label{s:P}

For a matrix $\bs z=(z_{ij})$ we use $\bs z^j$ to denote the $j$th column of $\bs z$, we use $\bs z^{j:k}$ to denote the submatrix formed from columns $j,\dotsc,k$, we use $\bs R(\bs z)=(R_i(\bs z))$ to denote the column vector of row sums defined by $R_i=\sum_j z_{ij}$, and we use $\bs C(\bs z)=(C_j(\bs z))$ to denote the row vector of column sums defined by $C_j=\sum_i z_{ij}$.  Fix the size of the matrix, $m\times n$, the weights $\bs w$, and the margins, $\bs r$ and $\bs c$, and let $\bs Z$ have distribution $P^*$ defined in \eqref{e} with $\Omega^*=\{\bs z\in\{0,1\}^{m\times n}:\bs R(\bs z)=\bs r,\ \bs C(\bs z)=\bs c\}$.  

To sample from $P^*$ we need only design a generic algorithm (generic in $m,n,\bs r,\bs c,\bs w$) for sampling from the distribution $P$ of the first column, namely, 
\[ P(\bs x)=\Prob(\bs Z^1=\bs x) . \]  The reason is that the conditional distribution of $\bs Z^{2:n}$ given $\bs Z^{1}$ has the same form as $P^*$ in \eqref{e}, but with different parameters.  The size of the matrix is now $m\times(n-1)$, the row sums are updated to $\bs r-\bs R(\bs Z^1)$, the column sums are updated to $\bs c^{2:n}$, and the weight matrix is updated to $\bs w^{2:n}$.  Once we have sampled $\bs Z^1$, then we can update these parameters and effectively start over, treating the second column of $\bs Z$ like it was the first column of the new, updated problem, and then continuing sequentially until we have sampled the entire matrix.  This is the same sequential strategy suggested by \citet{Chen:Sequential:2005}.  The supplementary material (located at the end of this document) contains a more detailed description of this column-wise factorization.

Henceforth, our target distribution is $P$, the distribution of $\bs Z^1$.  Let $\bs Y$ be a random matrix chosen uniformly over $\Omega^*$, let $\Omega$ denote the support of $\bs Y^1$, namely,
\[ \Omega = \{\bs x\in\{0,1\}^{m\times 1}:\bs x = \bs z^1, \bs z\in\Omega^*\} , \]
and for $\bs x\in\Omega$ define
\[ U(\bs x) = \Prob(\bs Y^1=\bs x) , \quad \quad V(\bs x)=\Exp\Bigl(\prod_{ij} w_{ij}^{Y_{ij}} \Bigl| \bs Y^1=\bs x\Bigr) . \]
 It is straightforward to verify that
\begin{equation}  P(\bs x) \propto U(\bs x)V(\bs x)\ind\{\bs x\in\Omega\} 
\label{e:Pparts}
\end{equation}
for $\bs x\in\{0,1\}^{m\times 1}$.
This factorization conceptually isolates the hard margin constraints, $\Omega$, the combinatorics, $U$, and the non-uniform weighting, $V$.  Although the separation is clearly artificial, it is useful to treat each of these factors separately when developing a proposal distribution. 

\subsection{The proposal distribution for the first column}

Motivated by the factorization in \eqref{e:Pparts}, we consider proposal distributions for the first column of the form
\[ Q(\bs x) \propto \tilde U(\bs x) \tilde V(\bs x) \ind\{\bs x\in\tilde\Omega\} , \]
where $\tilde U$ and $\tilde V$ are approximations of $U$ and $V$, respectively, that factor according to
\[ \tilde U(\bs x) \propto \prod_{i=1}^m u_i^{x_i} , \quad \quad \tilde V(\bs x) \propto \prod_{i=1}^m v_i^{x_i} \]
for some $\bs u,\bs v\in[0,\infty)^{m\times 1}$, and where $\tilde\Omega$ is of the form
\begin{equation} \label{e:Omega-tilde} \tilde\Omega = \bigl\{ \bs x \in\{0,1\}^{m\times 1} : x_{\pi_i}\in\mathcal{A}_i, \  \textstyle \sum_{\ell=1}^i x_{\pi_\ell}\in\mathcal{B}_i , \ i=1,\dotsc,m \bigr\} \end{equation}
for some permutation $\bs \pi=(\pi_1,\dotsc,\pi_m)$ of $(1,\dotsc,m)$ and some subsets $\bs{\mathcal{A}}=\mathcal{A}_1\times\dotsb\times\mathcal{A}_m\subseteq\{0,1\}^m$ and $\bs{\mathcal{B}}=\mathcal{B}_1\times\dotsb\times\mathcal{B}_m\subseteq\{0,1,\dotsc,c_1\}^m$.  Combining these approximations creates a proposal distribution of the form
\begin{equation} \label{e:Q}
 Q(\bs x) \propto \prod_{i=1}^m u_i^{x_i}v_i^{x_i}\ind\bigl\{x_{\pi_i}\in \mathcal{A}_i, \ \textstyle \sum_{\ell=1}^i x_{\pi_\ell} \in \mathcal{B}_i \bigr\} \quad\quad\quad (\bs x\in\{0,1\}^{m\times 1}) .  \end{equation}
Any proposal distribution of this form permits fast, exact sampling and evaluation using $O(mc_1)$ operations; see Section \ref{s:DP}.  The challenge is to find easily computable choices of $\bs u$, $\bs v$, $\bs \pi$, $\bs{\mathcal{A}}$, and $\bs{\mathcal{B}}$ such that $Q$ is a good approximation to the target $P$.  Fortunately, this seems to be possible in many cases; see Section \ref{s:components}.

For importance sampling to work, the support of $Q$, which is a subset of $\tilde\Omega$, must contain the support of $P$, which is a subset of $\Omega$.  When $\bs w$ has no zero entries, we require $\bs u$ and $\bs v$ to be positive and we engineer $\tilde\Omega$ to exactly coincide with $\Omega$ so that $P$ and $Q$ both have support $\Omega$; see Section \ref{s:margin}.  When $\bs w$ does have zero entries, we modify $\bs v$ and $\tilde\Omega$ to exclude certain elements in $\Omega$, but only elements that are not in the support of $P$.  This ensures that the support of $Q$ contains the support of $P$, but the supports may no longer be identical.   
In this case, if the importance sampling algorithm generates a column that is not in the support of $P$, then as it sequentially generates additional columns it will eventually try to create a $Q$ that is identically zero, indicating that no assignment of the current column simultaneously satisfies the margin constraints and has positive weight.  At this point the algorithm can assign an importance weight of zero and terminate.  Certain patterns of zero weights make the algorithm highly inefficient because the algorithm rarely terminates with a nonzero importance weight.  In the supplementary material we discuss alternative choices of $\bs v$ and $\tilde\Omega$ that are more efficient for certain patterns of particular interest, including the special case of zeros only on the diagonal.

\subsection{Efficient sampling and evaluation of the proposal} \label{s:DP}

Let $\bs X\in\{0,1\}^{m\times 1}$ have a distribution $Q$ that factors according to \eqref{e:Q} above for some $\bs u$, $\bs v$, $\bs\pi$, $\mathcal{A}$, and $\mathcal{B}$.  Define the permuted partial sums $\bs S\in\{0,\dotsc,c_1\}^{m\times 1}$ according to $S_i=\sum_{\ell=1}^i X_{\pi_\ell}$ for each $i$, and note that $\bs X$ and $\bs S$ are in bijective correspondence.  The distribution of $\bs S$ factors according to
\begin{equation} \Prob(\bs S=\bs s) \propto \prod_{i=1}^m h_i(s_{i-1},s_i) \label{e:S} \end{equation}
for $h_i(s_{i-1},s_i) = u_{\pi_i}^{s_i-s_{i-1}}v_{\pi_i}^{s_i-s_{i-1}}\ind\{s_i-s_{i-1}\in \mathcal{A}_i, \  s_i \in \mathcal{B}_i \}$, 
where here and below we define $S_0= s_0= 0$ for notational convenience.

The factorization in \eqref{e:S} implies that $\bs S$ is a Markov chain.  If we were given the standard Markov chain representation
\begin{equation}  \Prob(\bs S=\bs s) = \prod_{i=1}^m \Prob(S_i=s_i\bigl|S_{i-1}=s_{i-1}) \label{e:Qpi} ,\end{equation}
then generating a random observation of $\bs S$ would be trivial.  It is known that dynamic programming can be used to convert from Gibbs random field representations like \eqref{e:S} into Bayesian network representations like \eqref{e:Qpi}; see, e.g., \citet{Frey:Graphical:1998}.  The next theorem, which is straightforward to verify \citep[cf.][]{harrison2009rate}, summarizes dynamic programming in this context.  

\begin{theorem} \label{t:DP} Let $(S_0,S_1,\dotsc,S_m)$ be a sequence of random variables where each $S_i$ takes values in the finite set $D_i$ and where $D_0=\{0\}$.  Suppose there exists a sequence of functions $h_i:D_{i-1}\times D_i\mapsto[0,\infty)$ for $i=1,\dotsc,m$ such that the distribution of $(S_1,\dotsc,S_m)$ can be expressed as
\[  \Prob\bigl(S_1=s_1,\dotsc,S_m=s_m\bigr) \propto \prod_{i=1}^m h_i(s_{i-1},s_i) . \]   Recursively define $g_m(s_{m-1},s_m) = h_m(s_{m-1},s_m)$ and
\[ g_i(s_{i-1},s_i) = h_i(s_{i-1},s_i)\sum_{s_{i+1}\in D_{i+1}} g_{i+1}(s_i,s_{i+1}) \quad (i=1,\dotsc,m-1), \]
where each $g_i$ is defined over $D_{i-1}\times D_i$.  Then $S_0,\dotsc,S_m$ is a Markov chain and
\[ \Prob\bigl(S_i=s_i\bigl|S_{i-1}=s_{i-1}\bigr) = \frac{g_i(s_{i-1},s_i)}{\sum_{t\in D_{i}} g_i(s_{i-1},t)} \quad (i=1,\dotsc,m) . \]
\end{theorem}

In the present context, $S_i\in\mathcal{B}_i\subseteq\{0,\dotsc,c_1\}$ for each $i$, so the algorithm described in Theorem \ref{t:DP} for converting from \eqref{e:S} to \eqref{e:Qpi} requires at most $O(mc_1^2)$ operations.  In fact, since in the present situation we have $h_i(s_{i-1},s_i)=0$ for $s_i-s_{i-1}\not\in\{0,1\}$, implying the same for $g_i$, this yields an algorithm that requires $O(mc_1)$ operations.  Instead of representing all $(c_1+1)^2$ combinations of $(s_{i-1},s_i)$, we represent only the $2c_1+1$ feasible combinations.  Once the representation in \eqref{e:Qpi} is computed, generating a random observation $\bs X$ from $Q$ or evaluating $Q(\bs x)$ at any $\bs x$ takes $O(m)$ operations.  

\section{Specification of components} \label{s:components}

\subsection{Margin constraints} \label{s:margin}

Here we discuss the construction of $\tilde\Omega$. In particular, the next theorem shows how to ensure that $\tilde\Omega=\Omega$ for easily computable choices of $\bs\pi$, $\bs{\mathcal{A}}$, and $\bs{\mathcal{B}}$.
\begin{theorem} \citep{Chen:Sequential:2005} \label{t:binary} Assume $\Omega^*\neq\emptyset$. Choose $\bs\pi$ so that $r_{\pi_1}\geq \dotsb\geq r_{\pi_m}$.  For each $i=1,\dotsc,m$, define 
\[ \mathcal{A}_i = \begin{cases}  \{0\} & (r_{\pi_i}=0);
 \\ \{0,1\} & (0 < r_{\pi_i} < n); \\
 \{1\} & (r_{\pi_i}=n), \end{cases} \quad \quad \quad
\mathcal{B}_i = \begin{cases} \{\max\{0,b_i\},\dotsc,c_1\} & (i< m) ; \\
\{ c_1\} & (i = m) , \end{cases}
\]
for $b_i = \sum_{\ell=1}^i (r_{\pi_\ell} -  \sum_{j=2}^{n}\ind\{c_j \geq \ell\})$.
Define $\tilde\Omega$ according to \eqref{e:Omega-tilde}.  Then $\tilde\Omega=\Omega$.  
\end{theorem}

It is instructive to see how these choices of $\bs\pi$, $\bs{\mathcal{A}}$, and $\bs{\mathcal{B}}$ ensure that $\tilde\Omega\supseteq\Omega$, which is the primary requirement for importance sampling.  The Gale--Ryser conditions \citep{gale1957tfn,ryser1957cpm} state that there is a binary matrix with margins $\bs r\in\{0,\dotsc,n\}^{m\times 1}$ and $\bs c\in\{0,\dotsc,m\}^{1\times n}$ if and only if $\sum_i r_i=\sum_j c_j$ and 
\[  \sum_{\ell=1}^i r_{\pi_\ell} \leq \sum_{\ell=1}^i \sum_j \ind\{c_j\geq \ell\} \quad \text{for all $i=1,\dotsc,m-1$}, \]
where the permutation $\bs \pi$ is chosen so that $r_{\pi_1}\geq\dotsb\geq r_{\pi_m}$. 
It is straightforward to see that $\bs x \in\{0,1\}^{m\times 1}$ will be in $\Omega$ exactly when there is a way to fill out the remaining $n-1$ columns of the binary matrix that obey the updated margins after accounting for $\bs x$.  In other words, $\bs x\in\Omega$ exactly when $\bs r-\bs x$ and $\bs c^{2:n}$ satisfy the Gale--Ryser conditions for the margins of an $m\times(n-1)$ binary matrix.   

The set $\bs{\mathcal{A}}$ is chosen so that $\bs x\in\bs{\mathcal{A}}$ if and only if $\bs r-\bs x\in\{0,\dotsc,n-1\}^{m\times 1}$.  The set  $\mathcal{B}_m$ is chosen so that $\sum_{i=1}^m x_i\in\mathcal{B}_m$ if and only if $\sum_{i=1}^m  r_i-x_i = \sum_{j=2}^n  c_j$.  This is equivalent to enforcing the column sum $\sum_{i=1}^m x_i =  c_1$.  Choosing the permutation $\bs \phi$ so that $ r_{\phi_1}-x_{\phi_1}\geq\dotsb\geq  r_{\phi_m}-x_{\phi_m}$, the remaining Gale--Ryser conditions are that 
\begin{equation} \sum_{\ell=1}^i ( r_{\phi_\ell}-x_{\phi_\ell}) \leq \sum_{\ell=1}^i \sum_{j\geq 2} \ind\{ c_j\geq \ell\} \quad \text{for all $i=1,\dotsc,m-1$}, \label{e:Bphi} \end{equation}
which implies that
\begin{equation}  \sum_{\ell=1}^i ( r_{\pi_\ell}-x_{\pi_\ell}) \leq \sum_{\ell=1}^i \sum_{j\geq 2} \ind\{ c_j\geq \ell\} \quad \text{for all $i=1,\dotsc,m-1$}, \label{e:Bpi} \end{equation}
since the permutation $\bs \phi$ makes the left side as large as possible.  Solving \eqref{e:Bpi} for $\sum_{\ell=1}^i x_{\pi_\ell}$ gives the bounds encoded in the $\mathcal{B}_i$ and shows that $\tilde\Omega\supseteq\Omega$.

We cannot use the permutation $\bs \phi$ in the construction of $Q$ because $\bs \phi$ depends on $\bs x$, however, \citet{Chen:Sequential:2005} further prove that \eqref{e:Bphi} and \eqref{e:Bpi} are in fact equivalent, which means we are in the ideal situation where $\Omega=\tilde\Omega$.  (Although they made use of the factorization in Theorem \ref{t:binary}, their proposal distributions were not of the form in \eqref{e:Q}, except in the special case where $\Omega=\{\bs x\in\{0,1\}^{m\times 1}:\sum_i x_i=c_1\}$.)  Furthermore, \citet{chen2007cit} provides an extension of Theorem \ref{t:binary} for the case where, in addition to the margin constraints, $\Omega^*$ also enforces a fixed pattern of structural zeros for which there is at most one structural zero in each row and column.  This includes the important special case of adjacency matrices of directed graphs; see supplementary material.  

\subsection{Combinatorial approximations} \label{s:U}

Here we discuss approximation of $U$ by $\tilde U$.  Define
\[ N_{m,n}(\bs r,\bs c) = \bigl|\{\bs z\in\{0,1\}^{m\times n}:\bs{R}(\bs z)=\bs r, \ \bs C(\bs z)=\bs c\}\bigr| \]
to be the number of $m\times n$ binary matrices with row sums $\bs r$ and column sums $\bs c$.  We have
\[ U(\bs x) = \Prob(\bs Y^1=\bs x) = \frac{N_{m,n-1}(\bs r-\bs x,\bs c^{2:n})}{N_{m,n}(\bs r,\bs c)} , \]
(where as before, $\bs Y$ is uniform over $\Omega^*$)
and we desire an approximation of the form
\begin{equation} U(\bs x) = \frac{N_{m,n-1}(\bs r-\bs x,\bs c^{2:n})}{N_{m,n}(\bs r,\bs c)} \approx \gamma \prod_{i=1}^m u_i^{x_i} , \label{e:Nprod} \end{equation}
where $\gamma$ is an irrelevant positive constant.    


Temporarily pretending that \eqref{e:Nprod} is accurate for any $\bs x\in\{0,1\}^{m\times 1}$, we have
\[ u_i \approx \frac{U(\bs I^i)}{U(\bs 0)} =  \frac{N_{m,n-1}(\bs r-\bs I^i,\bs c^{2:n})}{N_{m,n-1}(\bs r,\bs c^{2:n})} , \]
where $\bs I^i$ is the $i$th column of the $m\times m$ identity matrix $\bs I$.  We cannot use this directly, since it is trying to evaluate $U$ outside of $\Omega$, and, furthermore, computationally efficient procedures for evaluating $N_{m,n-1}$ are not available.  Nevertheless, it suggests using
\begin{equation} u_i = \tilde N_{m,n-1}(\bs r-\bs I^i,\bs c^{2:n})/\tilde N_{m,n-1}(\bs r,\bs c^{2:n}) \quad \quad (i=1,\dotsc,m)  \label{e:u} \end{equation}
for an approximation $\tilde N$ of $N$ that extends smoothly to invalid margins.

Several asymptotic approximations for $N$ have appeared in the literature and could be used for $\tilde N$.  For example, \citet[Theorem 1]{canfield2008aed} suggest the following, which we write asymmetrically with respect to $\bs r$ and $\bs c$ in order to simplify \eqref{e:canfield-u} below:
\[ \begin{gathered}  \tilde N_{m,n}(\bs r,\bs c) = \binom{mn}{\sum_{k=1}^n c_k}^{-1}\prod_{i=1}^m \binom{n}{r_i}\prod_{j=1}^n \binom{m}{c_j} \exp\Bigl(-\frac{1}{2}\bigl(1-\mu_{m,n}(\bs r,\bs c)\bigr)\bigl(1-\nu_{m,n}(\bs c)\bigr)\Bigr) , \\
\mu_{m,n}(\bs r,\bs c) = \eta_{m,n}(\bs c)\sum_{i=1}^m \biggl(r_i-\frac{1}{m}\sum_{k=1}^n c_k\biggr)^2 , \quad \quad \nu_{m,n}(\bs c) = \eta_{m,n}(\bs c)\sum_{j=1}^n \biggl(c_j-\frac{1}{n}\sum_{k=1}^n c_k\biggr)^2  , \\
\eta_{m,n}(\bs c) = \frac{mn}{(\sum_{k=1}^n c_k) (mn-\sum_{k=1}^n c_k)} .
\end{gathered} \]
Substituting this into \eqref{e:u} and simplifying gives
\begin{equation} \label{e:canfield-u} u_i = \frac{r_i}{n-r_i}\exp\biggl[\eta_{m,n-1}(\bs c^{2:n})\bigl(1-\nu_{m,n-1}(\bs c^{2:n})\bigr)\biggl(\frac{1}{2}-r_i+\frac{1}{m}\sum_{k=2}^n c_k \biggr)\biggr] . \end{equation}
If $r_i=0$ or $r_i=n$, then the value of $X_i$ is determined by $\tilde\Omega$, and any choice of $u_i>0$ gives the same $Q$; we use $u_i=1$ in these cases.
We find that \eqref{e:canfield-u} works well over a large range of margins when $P^*$ is uniform.  It is excellent if the margins are approximately semi-regular, that is, if the row and column sums do not deviate substantially from their respective mean values.  

For certain pathological cases with wildly varying margins, such as those in \citet{bezakova2006nes}, \eqref{e:canfield-u} does not work well.  However, if the margins are such that the resulting matrices have a very low density of ones, even if the margins are highly irregular, then good performance can be obtained by instead using the asymptotic approximation of $N$ from \citet[Theorem 1.3]{greenhill2006aes}.  Details are provided in the supplementary material.  In fact, for the specific pathological cases in \citet{bezakova2006nes} using this alternative approximation gives $Q^*\equiv P^*$ in the uniform case.  None of the computationally efficient combinatorial approximations that we have found in literature work well when the margins are both highly irregular and lead to a moderate density of ones, but we are hopeful that advances in asymptotic enumeration techniques will eventually lead to approximations that work well in almost all cases.    

\citet{Chen:Sequential:2005} observed that combinatorial approximations could be used to find a good choice of $\bs u$ and they mentioned an early asymptotic approximation from \citet{Oneil:Asymptotics:1969}, which was explored further by \citet{blanchet2006isa} in an asymptotic analysis of the algorithm.  The examples in \citet{Chen:Sequential:2005}, however, use $u_i=r_i/(n-r_i)$, which is motivated by considering only the row margin constraints.  Although there are several substantial differences between their proposal distribution and ours for the special case of the uniform distribution of $\Omega^*$, we suspect that much of the improved performance of our algorithm results from using more accurate combinatorial approximations. 

In the next section we use $V$ and $\tilde V$ to account for the effects of $\bs w$, including the effects of zeros in $\bs w$.  Since these zeros affect the size of the support of $P^*$, an alternative, perhaps more natural approach is to allow $U$ and $\tilde U$ to capture the effects of zeros in $\bs w$.  The supplementary material contains more details.

\subsection{Non-uniform weighting} \label{s:V}

Here we discuss approximation of $V$ by $\tilde V$.  To develop an approximation, we will ignore the column margins and consider only the row margins.  This is similar to the approach used by \citet{Chen:Sequential:2005} to develop combinatorial approximations for the uniform case.  Let $\bs B\in\{0,1\}^{m\times n}$ be a matrix of independent Bernoulli$(1/2)$ random variables, so that
\begin{align} V(\bs x) & = \Exp\Bigl(\prod_{ij} w_{ij}^{Y_{ij}} \Bigl| \bs Y^1=\bs x\Bigr) = \Exp\Bigl(\prod_{ij} w_{ij}^{B_{ij}} \Bigl| \bs B^1=\bs x, \bs R(\bs B)=\bs r, \bs C(\bs B)=\bs c\Bigr) \notag 
\\
&  
\approx \Exp\Bigl(\prod_{ij} w_{ij}^{B_{ij}} \Bigl| \bs B^1=\bs x, \bs R(\bs B)=\bs r\Bigr) = \prod_{i=1}^m \Exp\Bigl(\prod_{j=1}^n w_{ij}^{B_{ij}} \Bigl| B_{i1}=x_i, R_i(\bs B)=r_i\Bigr) \notag 
\\
&  
\propto \prod_{i=1}^m v_i^{x_i} ,
\label{e:tildeV} \end{align}
where
\begin{align} v_i & = \frac{\Exp\bigl(\prod_{j=1}^n w_{ij}^{B_{ij}} \bigl| B_{i1}=1, R_i(\bs B)=r_i\bigr)}{\Exp\bigl(\prod_{j=1}^n w_{ij}^{B_{ij}} \bigl| B_{i1}=0, R_i(\bs B)=r_i\bigr)}  \quad \quad (r_i=1,\dotsc,n-1; \ i=1,\dotsc,m) \notag
\\
& = \frac{w_{i1}\binom{n-1}{r_i-1}^{-1}\sum_{\bs b\in\{0,1\}^{n-1}} \ind\{{\textstyle \sum_{j=1}^{n-1} b_j=r_i-1}\}\prod_{j=2}^n w_{ij}^{b_{j-1}}}{\binom{n-1}{r_i}^{-1}\sum_{\bs b\in\{0,1\}^{n-1}} \ind\{{\textstyle \sum_{j=1}^{n-1} b_j=r_i}\}\prod_{j=2}^n  w_{ij}^{b_{j-1}}} .
\label{e:v1} 
\end{align}
In the supplementary material we describe how to compute all possible $\bs v$ for all columns using $O(nd)$ operations (where $d=\sum_i r_i=\sum_j c_j$) in a one-time preprocessing step that can be done prior to sampling.  As with $u_i$, we always define $v_i=1$ if $r_i=0$ or $r_i=n$.  For cases where $\bs w$ has zeros, we can sometimes have a zero in the denominator of \eqref{e:v1} for $r_i>0$.  This happens when fewer than $r_i$ of the $n-1$ remaining weights in the row are nonzero.  Consequently, we need to force $X_i=1$, which we do by setting the corresponding $\mathcal A_{\ell}=\{1\}$ in Section \ref{s:margin}.

An important observation that we have thus far neglected is that many different choices of $\bs w$ give rise to the same $P^*$.  Define
\[ \Lambda(\bs w) = \{\bs t \in [0,\infty)^{m\times n} : \bs t = \bs \alpha\bs \beta^\tr\circ \bs w,\ \bs \alpha\in(0,\infty)^{m\times 1}, \ \bs \beta\in(0,\infty)^{n\times 1}\} , \]
where $\tr$ denotes transpose and $\bs y\circ \bs z$ is the Hadamard product, that is, element-wise multiplication of matrices of the same size, defined by $(\bs y\circ \bs z)_{ij}=y_{ij}z_{ij}$.
Then for every $\bs t\in \Lambda(\bs w)$ it is straightforward to verify that the $P^*$ defined with the weight matrix $\bs w$ and the $P^*$ defined with the weight matrix $\bs t$ are identical.  Similarly, the two versions of $V$ differ only by an inconsequential constant of proportionality.  Unfortunately, our approximation $\tilde V(\bs x)\propto\prod_i v_i^{x_i}$ defined above does not share this invariance.  Consequently, proposal distributions constructed with $\bs w$ and $\bs t$, respectively, could differ, even though the target distribution does not differ.  
We find this unappealing and remedy it in a preprocessing step prior to the construction of $Q^*$ by first transforming $\bs w$ into an equivalent, canonical $\bar{\bs w}\in\Lambda(\bs w)$.  In particular, $\bar{\bs w}$ is the unique element of $\Lambda(\bs w)$ with the property that its average nonzero entry over any row or column is one, namely,
\begin{equation} \sum_{i=1}^m \bar w_{ij} = \sum_{i=1}^m \ind\{\bar w_{ij} > 0\} ,\quad  \sum_{j=1}^n \bar w_{ij} = \sum_{j=1}^n \ind\{\bar w_{ij} > 0\}  \quad (i=1,\dotsc,m;\ j=1,\dotsc,n) . \label{e:barwH} \end{equation}
The solution to \eqref{e:barwH} over $\Lambda(\bs w)$ exists, is unique, and is easy to find numerically \citep{rothblum1989scalings}; see supplementary material for details and more discussion.  In the examples below, we always define $\bs v$ in terms of $\bar{\bs w}$, not $\bs w$.  Not only does this ensure that $Q^*$ has the same invariance property as $P^*$, but we find that performance of the algorithm tends to improve.

If we know that $P^*$ is uniform over $\Omega^*$, for example, if $\bs w\equiv 1$ or $\bar{\bs w}\equiv 1$, then $V$ is constant over $\Omega$, and we can ignore $\bs v$ in the construction of $Q$.

\section{Importance sampling}

\subsection{Algorithm summary} \label{s:overview}

\begin{enumerate}
\item  \label{l:pre} Preprocessing: Compute $\bar{\bs w}$ from \eqref{e:barwH} and precompute all possible $\bs v$ using \eqref{e:v1} with $\bar{\bs w}$ in place of $\bs w$; see supplementary material for details.  Compute $\bs\pi$, $\bs{\mathcal{A}}$, and $\bs{\mathcal{B}}$ according to Theorem \ref{t:binary} and $\bs{u}$ according to \eqref{e:canfield-u} for the first column.
\item \label{l:loop} Generating a single observation, $\bs{\tilde Z}$, from $Q^*$:  The matrix $\bs{\tilde Z}$ is generated column-by-column as follows. Set $q=1$. Sequentially, for each column:  
\begin{enumerate}
\item \label{2a} For the current $m,n,\bs r,\bs c$, compute $\bs\pi,\bs{\mathcal{A}},\bs{\mathcal{B}},\bs u$ as above.  After the preprocessing, updating these quantities based on the previously sampled column requires $O(m)$ operations.
\item Use Theorem \ref{t:DP} to compute the Markov chain representation for $Q$.  For the $j$th column, this takes $O(mc_j)$ operations.  If $Q\equiv 0$, then $\bs{\tilde Z}$ will not be in the support of $P^*$; assign a final importance weight of zero and go to step \ref{l:done}.
\item \label{l:makeX} Generate a random observation $\bs X$ from $Q$ and evaluate $Q(\bs X)$.  This takes $O(m)$ operations.  
\item Assign the current column of $\bs{\tilde Z}$ to be $\bs X$.  Update $q \leftarrow qQ(\bs X)$, $\bs c \leftarrow \bs c^{2:n}$, $n\leftarrow n-1$, $\bs r\leftarrow \bs r-\bs X$.  If $n >0$, continue looping over columns; go to step \ref{2a}.  If $n=0$, the final matrix is $\bs{\tilde Z}$, and we have $Q^*(\bs{\tilde Z})=q$; go to step \ref{l:done}.    
\end{enumerate}
\item \label{l:done} To generate additional independent observations from $Q^*$, reset all variables to their original values after step \ref{l:pre} and repeat step \ref{l:loop}.
\end{enumerate}
The same algorithm can be used to evaluate $Q^*(\bs z)$ for any $\bs z\in\Omega^*$.  Simply assign $\bs X$ to be the current column of $\bs z$ in step \ref{l:makeX}, instead of sampling a new column.  (The algorithm can be applied for any ordering of the columns, and $Q^*$ will depend on the chosen ordering.  The supplementary material describes the heuristics that we use to choose a column ordering.)

\subsection{Monte Carlo approximation and diagnostics}

Let $\bs Z$ have distribution $P^*$, let $\bs Z_1,\dotsc,\bs Z_T$ be random sample from $Q^*$ generated as above, and let $h$ be a function over $\Omega^*$.  Define the unnormalized importance weights
\[ f(\bs z) = \frac{\kappa P^*(\bs z)}{Q^*(\bs z)} = \frac{\prod_{ij} w_{ij}^{z_{ij}}\ind\{\bs z\in\Omega^*\}}{Q^*(\bs z)}  \quad \quad (\bs z\in\{0,1\}^{m\times n}) , \]
which we can efficiently evaluate for any $\bs z$ as described above.  In the formula for $f(\bs z)$ it is important that we use $\bs w$, not $\bar{\bs  w}$, particularly if we are approximating $\kappa$.  
We can approximate $\kappa$ and $\mu=\Exp(h(\bs Z))$ via importance sampling in the usual way, namely,
\begin{equation} \hat\mu_T = \frac{\sum_{t=1}^T f(\bs Z_t) h(\bs Z_t)}{\sum_{t=1}^T f(\bs Z_t)} \to \mu , \quad\quad \hat\kappa_T = \frac{1}{T}\sum_{t=1}^T f(\bs Z_t) \to \kappa \quad \quad (T\to\infty) .  \label{e:kappahat} \end{equation}
Besides being consistent, $\hat\kappa_T$ and $\hat\kappa_T\hat\mu_T$ are also unbiased approximations of $\kappa$ and $\kappa\mu$, respectively.  See \citet{Liu:Monte:2001} for details about importance sampling.  See \citet{harrison2012conservative} for modifications when \eqref{e:kappahat} is used to approximate a p-value.

In this context, importance sampling algorithms are usually evaluated empirically by diagnostics related to the variability of the importance weights.  The less variable the importance weights, the better the algorithm is judged to be performing.  For the numerical illustrations below, we report
\[ \hat\cv_T^2 = \frac{1}{(T-1)\hat\kappa_T^2}\sum_{t=1}^T (f(\bs Z_t)-\hat\kappa_T)^2 , \quad \quad \hat \Delta_T = \frac{\max_{t=1,\dotsc,T} f(\bs Z_t)}{\min_{t=1,\dotsc,T} f(\bs Z_t)}-1. \]
As $T\to\infty$, the approximate squared coefficient of variation, $\hat\cv_T^2$, converges to the true squared coefficient of variation, $\cv^2=\Var(f(\bs Z_1))/\Exp(f(\bs Z_1))^2=\Var(P^*(\bs Z_1)/Q^*(\bs Z_1))$, where $\Var$ denotes variance.  $\tilde T = T/(1+\cv^2)$ has been suggested as a rough diagnostic for effective sample size, meaning that a sample size of $T$ from $Q^*$ behaves roughly like a sample size of $\tilde T$ from $P^*$ for the purposes of Monte Carlo approximating $\mu$ for well-behaved functions $h$ \citep{Kong:Sequential:1994,Liu:Monte:2001}.  For many but not all examples we find $\hat\cv^2_T < 1$, suggesting that $Q^*$ is appropriate for efficient importance sampling.  The relative range of importance weights reported by $\hat \Delta_T$ is an especially stringent diagnostic.  For nearly constant margins and $P^*$ close to uniform, we often find $\hat \Delta_T\approx 0$, suggesting that $Q^*$ is an excellent approximation of $P^*$; see Table \ref{t:1}.

\section{Numerical illustrations} \label{s:sim}

We experiment with four different classes of weights based on a canonical matrix $\bs y$ whose entries are independently sampled from the uniform$(0,1)$ distribution: (I) $w_{ij}= 1$, which is the uniform distribution over $\Omega^*$, (II) $w_{ij}=y_{ij}+1$, (III) $w_{ij}=y_{ij}$, and (IV) $w_{ij}=-\ind(y_{ij}<0.99)\log(y_{ij})$, for all $i,j$. The specific entries of $\bs y$ for different sized matrices are in the supplementary material.  The resulting $P^*$ is increasingly non-uniform in each of the latter three cases and has $1\%$ structural zeros in case (IV).  Recall that each $\bs w$ corresponds to a family of weights of the form $\bs \alpha\bs \beta^\tr\circ \bs w$ that give the same $P^*$ and $Q^*$; see Section \ref{s:V}.  In all cases we report results with $T=1000$.

We begin with $500\times 500$ $r_1$-regular matrices, i.e., $r_i=c_j=r_1$ for all $i,j=1,\dotsc,500$.  Results are summarized in Table \ref{t:1} for $r_1=1,2,4,8,\dotsc,256$.  The diagnostics are striking, especially in the uniform case, for which the importance weights are essentially constant. 
Performance degrades slightly as $P^*$ becomes strongly non-uniform, but in all cases the estimated $\cv^2$ is less than one.  Low variability in importance weights corresponds to high precision in estimates of $\kappa$.  For example, in the uniform case, where $\kappa=|\Omega^*|$, for $r_1=256$ we obtain $\hat\kappa = (1.478301\pm0.000044){\times}10^{73781}$, where the errors are approximate standard errors estimated from the same importance samples,  and for $r_1=2$ we obtain $\hat\kappa_T = (2.27653\pm0.00017){\times}10^{2266}$, the latter of which is close to the true value of $\kappa=2.27658\dotsc{\times}10^{2266}$; see supplementary material.  To our knowledge the exact value of $\kappa$ in these examples can only be efficiently computed for the special case of the uniform distribution over either $1$-regular or $2$-regular matrices \citep{anand1966combinatorial}.  Sampling from the uniform distribution over $1$-regular matrices is trivial, $\kappa=m!$,  and there is no need to use our algorithm, although it is comforting that $Q^*=P^*$ in this case.  

We remark that the distributions corresponding to different weight classes in Table \ref{t:1} are almost singular with respect to each other.  For example, in the $1$-regular case, if we use the $Q^*$ for weight class I as a proposal distribution for the $P^*$ corresponding to one of the other weight classes, then we obtain, for weight class II, $\hat\Delta_T=2{\times}10^{13}$ and $\hat\cv_T^2=8{\times}10^2$, for weight class III, $\hat\Delta_T=2{\times}10^{67}$ and $\hat\cv_T^2=1{\times}10^3$, and for weight class IV, only $6$ of the $1000$ observations from $Q^*$ were even in the support of $P^*$, owing to the structural zeros.  Results are similar for other combinations and become even more extreme as $r_1$ increases.

\begin{table}[h]
\centering
\caption{$500\times 500$ $r_1$-regular matrices}
{\footnotesize
\begin{tabular}{ccccccccccccc}
& & \multicolumn{2}{c}{uniform} & & \multicolumn{2}{c}{$\bs w$ class II} & & \multicolumn{2}{c}{$\bs w$ class III} & & \multicolumn{2}{c}{$\bs w$ class IV} \\ [1ex]
$r_1$$\vphantom{\hat{\bigl(}}$ & & $\hat\Delta_T$ & $\hat\cv^2_T$ & & $\hat\Delta_T$ & $\hat\cv^2_T$ & & $\hat\Delta_T$ & $\hat\cv^2_T$ & & $\hat\Delta_T$ & $\hat\cv^2_T$ \\ [1ex] 
$1$  & &  $0$ &  $0$ & &  $2{\times}10^{-1}$ &  $5{\times}10^{-4}$ & &  $4{\times}10^{0}$ &  $4{\times}10^{-2}$ & &  $5{\times}10^{1}$ &  $3{\times}10^{-1}$ \\ 
 $2$  & &  $4{\times}10^{-2}$ &  $5{\times}10^{-6}$ & &  $2{\times}10^{-1}$ &  $4{\times}10^{-4}$ & &  $6{\times}10^{0}$ &  $4{\times}10^{-2}$ & &  $8{\times}10^{1}$ &  $2{\times}10^{-1}$ \\ 
 $4$  & &  $1{\times}10^{-2}$ &  $1{\times}10^{-6}$ & &  $1{\times}10^{-1}$ &  $4{\times}10^{-4}$ & &  $5{\times}10^{0}$ &  $3{\times}10^{-2}$ & &  $2{\times}10^{2}$ &  $2{\times}10^{-1}$ \\ 
 $8$  & &  $2{\times}10^{-2}$ &  $1{\times}10^{-6}$ & &  $2{\times}10^{-1}$ &  $3{\times}10^{-4}$ & &  $3{\times}10^{0}$ &  $3{\times}10^{-2}$ & &  $4{\times}10^{1}$ &  $2{\times}10^{-1}$ \\ 
 $16$  & &  $1{\times}10^{-2}$ &  $1{\times}10^{-6}$ & &  $2{\times}10^{-1}$ &  $3{\times}10^{-4}$ & &  $3{\times}10^{0}$ &  $3{\times}10^{-2}$ & &  $4{\times}10^{1}$ &  $1{\times}10^{-1}$ \\ 
 $32$  & &  $8{\times}10^{-3}$ &  $8{\times}10^{-7}$ & &  $1{\times}10^{-1}$ &  $2{\times}10^{-4}$ & &  $2{\times}10^{0}$ &  $2{\times}10^{-2}$ & &  $1{\times}10^{1}$ &  $1{\times}10^{-1}$ \\ 
 $64$  & &  $9{\times}10^{-3}$ &  $9{\times}10^{-7}$ & &  $1{\times}10^{-1}$ &  $2{\times}10^{-4}$ & &  $3{\times}10^{0}$ &  $2{\times}10^{-2}$ & &  $2{\times}10^{1}$ &  $9{\times}10^{-2}$ \\ 
 $128$  & &  $1{\times}10^{-2}$ &  $9{\times}10^{-7}$ & &  $1{\times}10^{-1}$ &  $9{\times}10^{-5}$ & &  $1{\times}10^{0}$ &  $1{\times}10^{-2}$ & &  $5{\times}10^{0}$ &  $5{\times}10^{-2}$ \\ 
 $256$  & &  $9{\times}10^{-3}$ &  $9{\times}10^{-7}$ & &  $5{\times}10^{-2}$ &  $5{\times}10^{-5}$ & &  $1{\times}10^{0}$ &  $1{\times}10^{-2}$ & &  $9{\times}10^{0}$ &  $7{\times}10^{-2}$
\end{tabular}
}
\label{t:1}
\end{table}

For the special case of $1$-regular matrices, corresponding to the first row in Table \ref{t:1}, $\kappa$ is the permanent of $\bs w$ and various generalizations of the permanent correspond to expectations under $P^*$.  The current state-of-the-art algorithm for approximating permanents and $\alpha$-permanents of general matrices, see \eqref{e:perm} above, seems to be the importance sampling algorithm of \citet{kou2009approximating}, which has about the same computational complexity as our algorithm.  For the case $\alpha=1$, their algorithm is nearly identical to ours, the main differences being the choice of column order and our use of $\bs{\bar w}$, which seems to give our algorithm slightly better performance.  Their algorithm is generally preferable for $\alpha\neq 1$, since it is tailored to the specific choice of $\alpha$, although in many cases performance is comparable.  The supplementary materials have numerical comparisons for each of the $\bs w$ used in Table \ref{t:1} and for all of the examples in \citet{kou2009approximating}, which include cases with $\alpha=1$ and $\alpha=1/2$.  It is interesting that in many cases our generic approach is competitive with specialized software.

In Table \ref{t:2} we repeat the simulations of Table \ref{t:1} for $50\times 100$ irregular matrices with margins $\bs r=k\tilde{\bs  r}$ and $\bs c=k\tilde{\bs  c}$, for the cases $k=1,\dotsc,4$, where ${\tilde{\bs  r}}^\tr= (24^1$, $22^2$, $17^4$, $13^3$, $12^2$, $11^3$, $10^2$, $9^3$, $8^6$, $7^1$, $6^4$, $5^4$, $4^5$, $3^6$, $2^4)$ and ${\tilde{\bs  c}}= (12^2$, $10^2$, $9^5$, $8^4$, $7^6$, $6^{11}$, $5^{10}$, $4^{18}$, $3^9$, $2^{13}$, $1^{20})$ using $i^j$ to denote $j$ copies of $i$.  Performance degrades in the irregular case as the matrices become more dense.  In most, but not all cases, the diagnostics suggest $Q^*$ could be used for efficient importance sampling.  

\begin{table}[h]
\centering
\caption{$50\times 100$ irregular matrices with $\bs r=k\tilde{\bs  r}$, $\bs c=k\tilde{\bs  c}$}
{\footnotesize
\begin{tabular}{ccccccccccccc}
& & \multicolumn{2}{c}{uniform} & & \multicolumn{2}{c}{$\bs w$ class II} & & \multicolumn{2}{c}{$\bs w$ class III} & & \multicolumn{2}{c}{$\bs w$ class IV} \\ [1ex]
$k$$\vphantom{\hat{\bigl(}}$ & & $\hat\Delta_T$ & $\hat\cv^2_T$ & & $\hat\Delta_T$ & $\hat\cv^2_T$ & & $\hat\Delta_T$ & $\hat\cv^2_T$ & & $\hat\Delta_T$ & $\hat\cv^2_T$ \\ [1ex] 
 $1$ 
  & &  $4{\times}10^{-1}$ &  $1{\times}10^{-3}$ 
  & &  $3{\times}10^{0\phantom{}}$ &  $5{\times}10^{-2}$ 
  & &  $8{\times}10^{1\phantom{}}$ &  $5{\times}10^{-1}$ 
  & &  $5{\times}10^{3\phantom{1}}$ &  $3{\times}10^{0\phantom{}}$ 
 \\ 
 $2$ 
  & &  $3{\times}10^{0\phantom{-}}$ &  $3{\times}10^{-2}$
  & &  $7{\times}10^{0\phantom{}}$ &  $1{\times}10^{-1}$ 
  & &  $7{\times}10^{2\phantom{}}$ &  $2{\times}10^{0\phantom{-}}$ 
  & &  $6{\times}10^{4\phantom{1}}$ &  $7{\times}10^{0\phantom{}}$ 
 \\ 
 $3$ 
  & &  $2{\times}10^{2\phantom{-}}$ &  $7{\times}10^{-1}$ 
  & &  $2{\times}10^{2\phantom{}}$ &  $6{\times}10^{-1}$ 
  & &  $2{\times}10^{4\phantom{}}$ &  $6{\times}10^{0\phantom{-}}$ 
  & &  $3{\times}10^{6\phantom{1}}$ &  $4{\times}10^{1\phantom{}}$ 
 \\ 
 $4$ 
  & &  $3{\times}10^{6\phantom{-}}$ &  $3{\times}10^{1\phantom{-}}$ 
  & &  $3{\times}10^{6\phantom{}}$ &  $2{\times}10^{1\phantom{-}}$ 
  & &  $4{\times}10^{9\phantom{}}$ &  $2{\times}10^{2\phantom{-}}$ 
  & &  $2{\times}10^{13}$ &  $8{\times}10^{2\phantom{}}$
\end{tabular}
}
\label{t:2}
\end{table}

For the special case of the uniform distribution, corresponding to the far left category of weights in Tables \ref{t:1} and \ref{t:2}, the sequential importance sampling algorithm of \citet{Chen:Sequential:2005}, as implemented in the publicly available R package {\tt networksis} \citep{admiraal2008networksis}, appears to be the current state-of-the-art algorithm for practical Monte Carlo approximation.  Our algorithm is a substantial improvement, especially for dense or irregular margins.  Using {\tt networksis} gives $\hat\Delta_T=(1{\times}10^1,1{\times}10^2,2{\times}10^5,1{\times}10^{11})$ and $\hat\cv^2_T=(2{\times}10^{-1},6{\times}10^{-1},1{\times}10^1,4{\times}10^2)$ for the first pair of columns in Table \ref{t:2}.  The {\tt networksis} implementation is several orders of magnitude slower than our implementation, and is too slow for most of the examples in Table \ref{t:1}.

\appendix

\section*{Supplementary  Material}

Supplementary material 
includes (i) a more detailed description of the column-wise factorization described in Section \ref{s:P}, (ii) details about the solution to \eqref{e:barwH} and other preprocessing of the weights and margins, (iii) alternative combinatorial approximations for sparse matrices with irregular margins, (iv) extensions to Theorem \ref{t:binary} for the case of structural zeros with at most one structural zero in each row and column, including the case of structural zeros along the diagonal, (v) more principled treatments of structural zeros in the approximations to $U$ and $V$, (vi) details for the numerical simulations, (vii) additional numerical illustrations, including examples using real data, and (viii) a Matlab implementation of the algorithm.

\section{Column-wise factorization}
 
Define the set of binary matrices with margins $\bs r\in\mathbb{N}^{m\times 1}$ and $\bs c\in\mathbb{N}^{1\times n}$ to be
\[ \Omega_{m,n}^*(\bs r,\bs c) = \{\bs z\in\{0,1\}^{m\times n} : \bs R(\bs z)=\bs r, \bs C(\bs z)=\bs c\} , \]
and let
\[ N_{m,n}(\bs r,\bs c) = |\Omega^*_{m,n}(\bs r,\bs c)| \]
denote the number of such matrices,
where $\mathbb{N}=\{0,1,\dotsc\}$ denotes the nonnegative integers.  For an $m\times n$ matrix $\bs w\in[0,\infty)^{m\times n}$ define the function
\[ P^*_{m,n}(\bs z\mid \bs r,\bs c,\bs w) = \frac{\ind\{\bs z\in\Omega^*_{m,n}(\bs r,\bs c)\}}{\kappa_{m,n}(\bs r,\bs c,\bs w)}\prod_{i=1}^m\prod_{j=1}^n w_{ij}^{z_{ij}} \quad \quad (\bs z\in\{0,1\}^{m\times n})  \]
with normalization constant
\[ \kappa_{m,n}(\bs r,\bs c,\bs w) = \sum_{\bs z\in\Omega^*_{m,n}(\bs r,\bs c)}\prod_{i=1}^m\prod_{j=1}^n w_{ij}^{z_{ij}}  \]
using the convention that $0/0=0$.  If $\kappa_{m,n}(\bs r,\bs c,\bs w)>0$, then $P^*_{m,n}$ is a probability mass function and we use $\bs Z$ to denote a random binary matrix with this distribution.  
We use $\bs Z^1$ to denote the first column of $\bs Z$, which has probability mass function
\[  P_{m,n}(\bs x\mid \bs r,\bs c,\bs w) = \Prob(\bs Z^1=\bs x) = \sum_{\bs z\in\Omega^*_{m,n}(\bs r,\bs c)} P^*_{m,n}(\bs z\mid \bs r,\bs c,\bs w)\ind\{\bs z^1=\bs x\} \quad \quad (\bs x\in\{0,1\}^{m\times 1}) , \]
the support of which is a subset of
\[ \Omega_{m,n}(\bs r,\bs c) = \{\bs x\in\{0,1\}^{m\times 1}:\bs x=\bs z^1,\ \bs z\in\Omega^*_{m,n}(\bs r,\bs c)\} . \]
If the entries of $\bs w$ are strictly positive, then the support is all of $\Omega_{m,n}(\bs r,\bs c)$.  

The algorithmic challenge of sampling the entire matrix $\bs Z$ reduces to the challenge of sampling from the first column $\bs Z^1$, because once we have the first column, then we can update the margins and proceed sequentially, treating successive columns like the first.  Indeed, it is straightforward to verify that
\[ \Prob(\bs Z^j=\bs x\mid \bs Z^{1:j-1}=\bs y) = P_{m,n-j+1}(\bs x \mid \bs r-\bs R(\bs y),\bs c^{j:n},{\bs  w}^{j:n}) , \]
so that the generic decomposition $\Prob(\bs Z) = \Prob(\bs Z^1)\prod_{j=2}^n \Prob(\bs Z^j\mid \bs Z^{1:j-1})$ gives
\[ P_{m,n}^*(\bs z\mid \bs r,\bs c,{\bs  w}) = P_{m,n}(\bs z^1\mid \bs r,\bs c,{\bs  w})\prod_{j=2}^n P_{m,n-j+1}(\bs z^j \mid \bs r-\bs R(\bs z^{1:j-1}),\bs c^{j:n},{\bs  w}^{j:n}) . \]
(In the main text, we primarily focus on sampling the first column $\bs Z_1$, we suppress $m,n,\bs r,\bs c,{\bs  w}$ in the notation as much as possible, and we assume that $\kappa > 0$.)  To summarize, our target distribution is the binary random vector $\bs Z_1\in\{0,1\}^{m\times 1}$ with probability mass function
\[ \label{e:P} P(\bs x) \propto \sum_{\bs z\in\{0,1\}^{m\times n}}\ind\{\bs R(\bs z)=\bs r,\bs C(\bs z)=\bs c,\bs z^1=\bs x\} \prod_{i=1}^m\prod_{j=1}^n w_{ij}^{z_{ij}} . \]

\section{Preprocessing the weights and margins} 

The preprocessing that affects the definition of $Q^*$ consists of transforming $\bs w$ into $\bar{\bs w}$ and choosing an ordering of the columns.  Other elements of the preprocessing are merely for computational efficiency.  All of the preprocessing of $\bs w$, but not reordering the columns, can be skipped when it is known that $P^*$ is uniform over $\Omega^*$, e.g., when $\bar{\bs  w}\equiv 1$.

\subsection{Computing $\bar{\bs{w}}$, the solution to equation (16) in the main text}

Fix $\bs w\in[0,\infty)^{m\times n}$.  Define 
\[ n_i = \sum_{j=1}^n \ind\{w_{ij} > 0\}, \quad m_j = \sum_{i=1}^m \ind\{w_{ij} > 0\} \quad (i=1,\dotsc,m; \ j=1,\dotsc,n) .\]
We are looking for the $\bar{\bs w}\in[0,\infty)^{m\times n}$ with the following properties:
\[
\bar w_{ij}=\alpha_i\beta_j w_{ij}, \quad \sum_{j=1}^n \bar w_{ij} = n_i, \quad \sum_{i=1}^m \bar w_{ij} = m_j \quad (\alpha_i,\beta_j >0; \ i = 1,\dots,m; \ j = 1,\dots,n) .
\]
Initializing $\bar{\bs w}^{(0)} = \bs w$ and $t=1$, we iterate the following fixed point equations until convergence:
\[ \begin{aligned} \bar w^{(2t-1)}_{ij} & = \frac{n_i\bar w^{(2t-2)}_{ij}}{\sum_{\ell=1}^n\bar w^{(2t-2)}_{i\ell}} \quad (i=1,\dotsc,m;\ j=1,\dotsc,n) \\
\bar w^{(2t)}_{ij} & = \frac{m_j\bar w^{(2t-1)}_{ij}}{\sum_{\ell=1}^m\bar w^{(2t-1)}_{\ell j}} \quad (i=1,\dotsc,m;\ j=1,\dotsc,n) ,
\end{aligned} \]
where superscripts are indices, and where we take $0/0=0$.  If we iterate this for $T$ steps, then we use $\bar{\bs w} = \bar{\bs w}^{(2T)}$.  In our experience, a small $T$ is usually adequate to reach convergence.  Since each $\bar{\bs w}^{(T)}\in\Lambda(\bs w)$, iterating to convergence is not important for validity of the algorithm.   \citet{rothblum1989scalings} prove existence and uniqueness of $\bar{\bs w}$.  They also show that the solution can also be found using a convex programming algorithm, but we have not experimented with this approach.

The computational cost of computing $\bar{\bs w}$ takes at least $O(mn)$ operations, but we do not have theoretical bounds on the computational complexity.  In our experience, it can be treated as a negligible preprocessing step.  Although this choice of $\bar{\bs w}$ outperforms many alternatives, we have found no theoretical justification for its use.  It is closely related to Sinkhorn balancing of $\bs w$ \citep{sinkhorn1964relationship,sinkhorn1967diagonal}, which has appeared in the literature in both algorithmic and theoretical treatments of permanents \citep[e.g.,][]{ando1989majorization,beichl1999approximating}, and it has the nice property that $\bar{\bs w}\equiv 1$ whenever $\bs w=\bs \alpha\bs \beta^\tr$.  In any case, $P^*_{m,n}(\cdot\mid \bs r,\bs c,\bar{\bs w})=P^*_{m,n}(\cdot\mid \bs r,\bs c,\bs w)$, so switching from $\bs w$ to $\bar{\bs w}$ does not change the target distribution.

\subsection{Choosing a column ordering}

Our algorithm is not invariant to the ordering of the columns, nor to the pattern of zeros in $\bar{\bs w}$.  We use the following heuristic ordering of the columns.  First, if $\bar{\bs w}$ is banded, then we leave the columns in their original order.  The special case of banded weights arises frequently in some applications and we find that the banded ordering works best for accommodating so many zero weights.  In other cases, we reorder the columns first by decreasing column sum and then by decreasing variance of the entries of $\bar{\bs  w}$ within each column.  These preprocessing steps, and the accompanying postprocessing steps of returning the columns to their original orders, all require negligible additional computation.  In practice, if one is interested in a specific matrix for which these heuristics do not work well, then it can often be advantageous to experiment with different column orders or perhaps swapping the roles of rows and columns.  For the description of the algorithm, when referring to the $j$th column, we mean the $j$th column after any reordering of the columns.

\subsection{Precomputing the constants $\bs v$ for all columns} \label{ss:v}

Define the symmetric polynomials 
\begin{equation} \label{e:Gn} G_{n}(\bs y,k) = \sum_{\bs b\in\{0,1\}^n} \ind\{{\textstyle\sum_{j=1}^n b_j=k}\}\prod_{j=1}^n y_j^{b_j} \quad \quad (\bs y\in\mathbb{R}^n, \ k=0,\dotsc,n) , \end{equation}
and let $\bar{\bs  w}_i^{j:k}=(\bar w_{ij},\dotsc,\bar w_{ik})$ denote the $i$th row of $\bar{\bs  w}^{j:k}$.
Before sampling we also precompute and store
\begin{equation} G(i,j,k) = G_{n-j+1}(\bar{\bs  w}_i^{j:n},k)  \label{e:G} \end{equation}
for all $i=1,\dotsc,m$, $j=1,\dotsc,n$, and $k=0,\dotsc,\min(r_i,n-j+1)$.  The entire collection can be computed in $O(nd)$ operations (where $d=\sum_i r_i =\sum_j c_j$), by initializing with $G(i,j,0) = 1$, $G(i,n,1)=\bar w_{in}$, and $G(i,j,k)=0$ for all $i,j$ and $k>n-j+1$, and then using the recursive formula
\[ G(i,j,k) = G(i,j+1,k)+\bar w_{ij} G(i,j+1,k-1) . \]
In particular, in equation (15) in the main text we see that for the first column
\[ v_i = \frac{\bar w_{i1}\binom{n-1}{r_i-1}^{-1} G(i,2,r_i-1)}{\binom{n-1}{r_i}^{-1}G(i,2,r_i)} . \]
(Recall that we use $\bar {\bs w}$, not $\bs w$, in our implementation of the algorithm.)
For the $j$th column ($1 < j < n$) we will have
\[ v_i = \frac{\bar w_{ij}\binom{n-j}{r_i-R_i(\bs z^{1:j-1})-1}^{-1} G(i,j+1,r_i-R_i(\bs z^{1:j-1})-1)}{\binom{n-j}{r_i-R_i(\bs z^{1:j-1})}^{-1}G(i,j+1,r_i-R_i(\bs z^{1:j-1}))} , \]
where $\bs z^{1:j-1}$ are the previously sampled columns so that $\bs r-\bs R(\bs z^{1:j-1})$ are the updated row sums when preparing to sample the $j$th column.

\section{Alternative combinatorial approximations} \label{s:c}

For each positive integer $\ell$ and any nonnegative integer $a$ we define 
\[ [a]_\ell = a(a-1)\dotsm(a-\ell+1) , \] and for a $k$-vector $\bs t$ of nonnegative integers we define
\[ [\bs t]_\ell = \sum_{i=1}^k [t_i]_\ell . \]

In Section 4.2 of the main text we used a combinatorial approximation due to \citet{canfield2008aed}, however, other approximations can also be used and may give better performance for some problems.  For instance, \citet[Theorem 1$.$3]{greenhill2006aes} provide an alternative combinatorial approximation for $N_{m,n}(\bs r,\bs c)$ that is accurate, asymptotically, for sparse matrices, except perhaps when the margins are extremely variable:
\begin{gather*}
\tilde N_{m,n}(\bs r,\bs c) = \frac{[\bs c]_1!}{\prod_{i=1}^m r_i!\prod_{j=1}^nc_j!}\exp\bigl(-\alpha_1(\bs{c})[\bs{r}]_2-\alpha_2(\bs{c})[\bs{r}]_3-\alpha_3(\bs{c})[\bs{r}]_2^2\bigr) ,\\
\alpha_1(\bs{c}) = \frac{[\bs{c}]_2}{2[\bs c]_1^2}+\frac{[\bs{c}]_2}{2[\bs c]_1^3}+\frac{[\bs{c}]_2^2}{4[\bs c]_1^4} , \quad \alpha_2(\bs{c}) = -\frac{[\bs{c}]_3}{3[\bs c]_1^3}+\frac{[\bs{c}]_2^2}{2[\bs c]_1^4} , \quad 
\alpha_3(\bs{c}) = \frac{[\bs{c}]_2}{4[\bs c]_1^4}+\frac{[\bs{c}]_3}{2[\bs c]_1^4}-\frac{[\bs{c}]_2^2}{2[\bs c]_1^5} , 
\end{gather*}
where we take $0/0=0$.  Following Section 4.2 in the main text, a straightforward calculation gives
\[ u_i = r_i\exp\bigl[(r_i-1)\bigl(2\alpha_1(\bs c^{2:n})+3\alpha_2(\bs c^{2:n})(r_i-2)+4\alpha_3(\bs c^{2:n})([\bs r]_2-r_i+1)\bigr)\bigr] . \]
This combinatorial approximation is an improvement of an approximation in \cite{Oneil:Asymptotics:1969}, which was mentioned in \citet{Chen:Sequential:2005} and studied by \citet{blanchet2006isa}.  Both are exactly uniform for the pathological cases in \cite{bezakova2006nes}; see Supplementary Section \ref{s:uniform}.

\section{Structural zeros and ones}

\subsection{Remarks}

The algorithm can be improved to better accommodate structural zeros.  We avoided this in the main text to simplify the exposition, but the complexity of the algorithm does not change significantly.  The numerical experiments in the main text do not use these improvements, even though some of the examples have structural zeros.

We use the term structural zeros to denote positions $(i,j)$ such that $w_{ij}=0$, which allows the investigator to explicitly force the binary matrix to zeros at those positions.  It is possible that the row and column sums also force some entries to be zero, but we are not referring to those types of implicit structural zeros.  

Sometimes it is desirable to force an entry to be one.  These structural ones can be accommodated using structural zeros.  In a preprocessing step, we replace structural ones with structural zeros and decrement the row and column sums appropriately.  Then we sample as usual.  In a postprocessing step, we reinsert the structural ones.  Henceforth, we only discuss structural zeros.

\subsection{Extensions to Theorem 2 in the main text for zero diagonal}

Here we report an extension of Theorem 2 in the main text to the case where $w$ has at most one zero entry in each row and column.  This includes the special case of a zero diagonal, which arises frequently when the binary matrices of interest are adjacency matrices of directed graphs.  Unlike the main text, the order of the columns is important for the validity of the algorithm.  The columns must be reordered during preprocessing so that $c_1\geq\dotsb\geq c_n$.
\begin{theorem} \citep{chen2006sec,chen2007cit} Assume that $c_1\geq\dotsb\geq c_n$, fix $\bs w\in[0,\infty)^{m\times n}$, define $a_{ij}=\ind\{w_{ij} > 0\}$ for each $i,j$, assume $R_i(\bs a)\geq n-1$ and $C_j(\bs a)\geq m-1$ for each $i,j$, and assume that $\kappa_{m,n}(\bs r,\bs c,\bs w) > 0$.  Choose $\bs\pi$ so that $r_{\pi_1}\geq\dotsb\geq r_{\pi_m}$ and so that whenever $r_{\pi_i}=r_{\pi_{i+1}}$ we also have $y_{\pi_i}\leq y_{\pi_{i+1}}$, where
\[ y_i = \begin{cases} \text{the unique $j$ such that $w_{ij}=0$} & \text{if there exists such a $j$;} \\ n+1 & \text{otherwise.} \end{cases} \]
For each $i=1,\dotsc,m$, define 
\[ \mathcal{A}_i = \begin{cases} \{0\} & (a_{\pi_{i}1}r_{\pi_i}=0); \\ \{0,1\} & (0 < a_{\pi_{i}1}r_{\pi_i} < R_{\pi_i}(a)); \\
\{1\} & (a_{\pi_{i}1}r_{\pi_i} = R_{\pi_i}(a)), \end{cases}
\quad \quad 
\mathcal{B}_i = \begin{cases} \{\max(0,b_i) , \dotsc, c_1\} & (i<m); \\
\{c_1\} & (i = m), \end{cases}
\]
for 
\[ b_i = \textstyle (\sum_{\ell=1}^i r_{\pi_\ell}) - \min_{j=1,\dotsc,n} \{ \sum_{k=j+1}^n c_k  + \sum_{\ell=1}^i\sum_{k=2}^j a_{\pi_\ell k}\} . \]
Define $\tilde\Omega$ according to (5) in the main text.  Then $\tilde\Omega$ is the support of $P_{m,n}(\cdot\mid \bs r,\bs c,\bs w)$. 
\end{theorem}

\subsection{Alternative treatments of structural zeros}

Here we redefine $\bs{\mathcal{A}}$, $U$, and $V$ from the main text to account for structural zeros differently.  We define $\bs{\mathcal{A}}$ according to Supplementary Theorem 3 above, which allows trivial cases to be handled by $\tilde\Omega$.

Let $\bs Y$ be a random matrix chosen uniformly over the support of $P^*$ and define
\[ U(\bs x) = \Prob(\bs Y^1=\bs x)=\frac{N_{m,n-1}(\bs r-\bs x,\bs c^{2:n})}{N_{m,n}(\bs r,\bs c)} , \quad \quad  V(\bs x)=\Exp\Bigl(\prod_{ij} w_{ij}^{Y_{ij}} \Bigl| \bs Y^1=\bs x\Bigr) , \]
so that, for any $\bs x$, $P(\bs x) \propto U(\bs x)V(\bs x)\ind\{\bs x\in\Omega\}$.  In the main text, these definition were the same except that $\bs Y$ was chosen uniformly over $\Omega^*$.  If $\bs w$ forces structural zeros, then the support of $P^*$ may be smaller than $\Omega^*$.  We proceed exactly as in the main text to develop approximations of the new $U$ and $V$.

For the new $U$, we follow section 4.2 and note that $N_{m,n}(\bs r,\bs c)$ needs to be replaced by the size of the support of $P^*_{m,n}(\cdot\mid \bs r,\bs c,\bs w)$, say $N_{m,n}(\bs r,\bs c,\bs w)$, and, consequently, $\tilde N$ needs to be replaced by a combinatorial approximation of the size of the support of $P^*$.  \citet{greenhill2009random} provide the modified asymptotic enumeration results corresponding to those that led to equation (13) in the main text.  Define $a_{ij}=\ind\{w_{ij} > 0\}$ for all $i,j$.  Note that $N_{m,n}(\bs r,\bs c,\bs w)=N_{m,n}(\bs r,\bs c,\bs a)$.  For an approximation of $\tilde N_{m,n}(\bs r,\bs c,\bs w)$, \citet[Theorem 2.1]{greenhill2009random} suggest
\[
\begin{gathered}
\begin{aligned}
\tilde N_{m,n}(\bs r,\bs c,\bs a) = & \binom{\sum_{\ell=1}^m\sum_{k=1}^n a_{\ell k}}{\sum_{k=1}^n c_k}^{-1}\prod_{i=1}^m \binom{R_i(\bs a)}{r_i}\prod_{j=1}^n \binom{C_j(\bs a)}{c_j} \\ & \times \exp\Bigl[-\frac{1}{2}\bigl(1-\mu_{m,n}(\bs r,\bs c)\bigr)\bigl(1-\nu_{m,n}(\bs c)\bigr)-\delta_{m,n}(\bs r,\bs c,\bs a)\Bigr] ,
\end{aligned} \\
\delta_{m,n}(\bs r,\bs c,\bs a) = \eta_{m,n}(\bs c)\sum_{i=1}^m\sum_{j=1}^n \left((1-a_{ij})\biggl(r_i-\frac{R_i(\bs a)}{mn}\sum_{k=1}^n c_k\biggr)\biggl(c_j-\frac{C_j(\bs a)}{mn}\sum_{k=1}^n c_k\biggr)\right) , \notag
\end{gathered}
\]
which reduces to the formula in the main text when $\bs a\equiv 1$.  The functions $\mu,\nu,\eta$ are defined in the main text.  This approximation leads to
\[ \begin{aligned} u_i = & \frac{r_i}{R_i(\bs a^{2:n})-r_i+1}\exp\Biggl(\eta_{m,n-1}(\bs c^{2:n})\biggl[\bigl(1-\nu_{m,n-1}(\bs c^{2:n})\bigr)\biggl(\frac{1}{2}-r_i+\frac{1}{m}\sum_{k=2}^n c_k\biggr) \\
& + \sum_{j=2}^n(1-a_{ij})\biggl(c_j-\frac{C_j(\bs a)}{m(n-1)}\sum_{k=2}^n c_k\biggr)\biggr]\Biggr) , \end{aligned} \]
where, as before, we set $u_i=1$ whenever $r_i=0$ or $r_i=R_i(\bs a^{2:n})+1$.

For the new $V$, we follow section 4.3 in the main text, but define $\bs B$ to be a matrix of independent Bernoulli random variables where $B_{ij}$ is Bernoulli$(a_{ij}/2)$ for $a_{ij}=\ind\{w_{ij} > 0\}$.  Following equations (14) and (15) from the main text, the first change comes after the second equality in (15), giving
\[ \begin{aligned} v_i & = \frac{w_{i1}\binom{R_i(\bs a^{2:n})}{r_i-1}^{-1}\sum_{\bs b\in\{0,1\}^{n-1}} \ind\{{\textstyle \sum_{j=1}^{n-1} b_j=r_i-1}\}\prod_{j=2}^n w_{ij}^{b_{j-1}}}{\binom{R_i(\bs a^{2:n})}{r_i}^{-1}\sum_{\bs b\in\{0,1\}^{n-1}} \ind\{{\textstyle \sum_{j=1}^{n-1} b_j=r_i}\}\prod_{j=2}^n w_{ij}^{b_{j-1}}}
\\ 
& =
\frac{w_{i1}(R_i(\bs a^{2:n})-r_i+1)G_{n-1}(\bs w^{2:n}_i,r_i-1)}{r_i G_{n-1}(\bs w^{2:n}_i,r_i)}
= \frac{w_{i1}(R_i(\bs a^{2:n})-r_i+1)G(i,2,r_i-1)}{r_i G(i,2,r_i)} , 
\end{aligned}
\]
where we have made use of the fact that terms inside the summations in the first expression are zero whenever $\bs b$ has an entry of one in a place where there is a structural zero.  The functions $G_n$ and $G$ are defined above in supplementary section \ref{ss:v}.  For zeros in the numerator or denominator of the expression for $v_i$ we set $v_i=1$ and allow $\mathcal{A}_i$ to deterministically choose the appropriate value of the $i$th entry.  As in the main text, we suggest replacing $\bs w$ with $\bar{\bs w}$ throughout.  

\section{Numerical illustrations}

\subsection{Pseudorandom number generator}

The pseudorandom number generator used by the importance sampling algorithm for the numerical illustrations is the default pseudorandom number generator in Matlab version 7.14, which is the Mersenne twister algorithm {\tt mt19937ar} \citep[c.f.][]{matsumoto1998mersenne} described at \\ {\tt http://www.math.sci.hiroshima-u.ac.jp/$\sim$m-mat/MT/emt.html} .

\subsection{Canonical weight matrices}

The weight matrices $\bs w$ used in Section 6 of the main text are built from a canonical matrix $\bs y$.  The $m\times n$ canonical matrix $\bs y$ is constructed as follows:
\[ y_{ij} = \frac{R((j-1)m+i)}{2^{31}-1}  \quad \quad (i=1,\dotsc,m; \ j=1,\dotsc,n) , \]
where $R(0)=1$ and
\[ R(k) = 7^5R(k-1)\bmod (2^{31}-1) \quad \quad (k=1,\dotsc,mn). \]
The sequence $R(1),R(2),\dotsc$ is a simple, well-known multiplicative congruential pseudorandom number generator, known as MINSTD, for the discrete uniform distribution over $\{1,\dotsc,2^{31}-2\}$ \citep{park1988random}.  It was the default pseudorandom number generator in Matlab for many years and is fine for our purpose of creating a matrix $\bs y$ with independent uniform$(0,1)$ entries whose values are easy to communicate to others.

\subsection{The number of $n\times n$ two-regular binary matrices} \label{s:2reg}

\citet[Eq.~(27)]{anand1966combinatorial} give a simple recursive formula for the number of $n\times n$ two-regular binary matrices, say $H_n$.  Initialize $H_1=0$, $H_2=1$, $H_3=6$, and then
\[ H_k = \frac{1}{2}k(k-1)^2\bigl((2k-3)H_{k-2}+(k-2)^2H_{k-3}\bigr) \quad \quad \quad (k=4,5,\dotsc) . \]
The exact value of $H_{500}$ can be found in the appendix of this supplement.  As noted in Section 6 of the main text, our algorithm provides an extremely accurate approximation.  

\cite{Chen:Sequential:2005} used their importance sampling algorithm to approximate $H_{100}$ as $(2.96\pm 0.03){\times}10^{314}$ based on a sample size of $100$.  For comparison, using a sample of size $100$ from our algorithm gives an approximation of $(2.969 \pm 0.001){\times} 10^{314}$, which appears to be almost $1000$ times more efficient for the purposes of approximate enumeration.  The true value is $2.9692\dotsc{\times}10^{314}$.  The full number can be found in the appendix of this supplement.  We should also note that the importance sampling approximations are much more accurate than the combinatorial approximations upon which the importance sampling algorithm is based.  For instance, using the approximation $\tilde N$ from Section 4.2 of the main text gives $2.957{\times}10^{314}$.

\subsection{Approximating $\alpha$-permanents}

Here we report comparisons between using our algorithm for approximating $\alpha$-permanents and using the custom importance sampling algorithm of \citet{kou2009approximating}.  We thank Sam Kou for sharing his code with us.  The $\alpha$-permanent of $\bs w$ can be expressed as
\[ \text{per}_\alpha(\bs w) = \kappa \Exp(\alpha^{\text{cyc}(\bs Z)}) , \]
where $\bs Z$ has distribution $P^*$ with the same $\bs w$ and all row and column sums equal to one; see equation (3) in the main text.  We approximate it using the consistent, unbiased approximation 
\[ \hat{\text{per}}_{\alpha,T}(\bs w) = \hat\kappa_T\hat\mu_T = \frac{1}{T}\sum_{t=1}^T f(\bs Z_t)h(\bs Z_t) \]
for $h(\bs z)=\alpha^{\text{cyc}(\bs z)}$; see Section 5.2 and equation (17) in the main text.

The Kou \& McCullagh algorithm does not attempt to generate $\bs Z$ from a distribution that is close to $P^*$, like ours does, but rather from a distribution proportional to $h(\bs z)P^*(\bs z)$.  In the case where $\alpha=1$ so that $h\equiv 1$, the two approaches agree and the empirical results are quite similar.  But when $\alpha\neq 1$, their algorithm is generally better, because is it tailored for the choice of $\alpha$.  Nevertheless, our algorithm might be useful in cases where $\text{per}_\alpha(\bs w)$ is needed for many $\alpha$ simultaneously, or in cases where $\alpha$ is very close to one.

Supplementary Table \ref{t:s1} reports $\hat{\text{per}}_{\alpha,T}(\bs w)$ along with approximate standard errors defined as $\hat\sigma_T/\sqrt{T}$, where
\[ \hat\sigma_T^2 = \frac{1}{T-1}\sum_{t=1}^T \bigl(f(Z_t)h(Z_t)-\hat\kappa_T\hat\mu_T\bigr)^2 . \]
It also reports an approximate relative standard error defined as
\[ \hat{\text{rel}}_T = \frac{\hat\sigma_T/\sqrt{T}}{\hat\kappa_T\hat\mu_T}\times 100\% . \]
We use $T=1000$ for the examples with $n=500$ to match Table 1 in the main text.  The other $\bs w$ are taken from \citet{kou2009approximating} and we use $T=20000$ to facilitate comparison with their results.  In some cases the true value of $\text{per}_\alpha(\bs w)$ is known and this is shown in the final column of the table; see the supplementary appendix.  Except for the $n= 500$ examples and the results from our algorithm, the entries of Supplementary Table \ref{t:s1} come directly from Table 1 in \citet{kou2009approximating}.  

\begin{table}[h]
\centering
\caption{Approximating $\alpha$-permanents}
{\small
\begin{tabular}{c@{~~}c@{~~}cc@{~~~}cc@{~~~}c@{~~~}c}
 \multicolumn{3}{c}{parameters} &  \multicolumn{2}{c}{our algorithm} & \multicolumn{2}{c}{Kou \& McCullagh} & true value \\ 
$\bs w$ & $n$ & $\alpha$ & $\hat{\text{per}}_{\alpha,T}(\bs w)$ & $\hat{\text{rel}}_T\%$ & $\hat{\text{per}}_{\alpha,T}(\bs w)$ & $\hat{\text{rel}}_T\%$ & $\text{per}_\alpha(\bs w)$ \\
I & $500$ & $1$ & $(1.220\pm 0.000){\times}10^{1134}$ & $0.00$ & $(1.220\pm 0.000){\times}10^{1134}$ & $0.00$ & $1.220{\times}10^{1134}$ \\
II & $500$ & $1$ & $(1.437 \pm 0.001){\times}10^{1222}$ & $0.08$ & $(1.441\pm 0.001){\times}10^{1222}$ & $0.08$ & ?  \\
III & $500$ & $1$ & $(3.998 \pm 0.028){\times}10^{983\phantom{0}}$ & $0.69$ & $(3.975 \pm 0.033){\times}10^{983\phantom{0}}$ & $0.82$ & ? \\
IV & $500$ & $1$ & $(3.523 \pm 0.056){\times}10^{1133}$ & $1.60$ & $(3.546 \pm 0.066){\times}10^{1133}$ & $1.85$ & ? \\
I & $500$ & $1/2$ & $(2.963 \pm 0.167){\times}10^{1132}$ & $5.62$ & $(3.078 \pm 0.000){\times}10^{1132}$ & $0.00$ & $3.078{\times}10^{1132}$ \\
II & $500$ & $1/2$ & $(3.296 \pm 0.174){\times}10^{1220}$ & $5.28$ & $(3.662 \pm 0.012){\times}10^{1220}$ & $0.32$ & ?\\
III & $500$ & $1/2$ & $(8.889 \pm 0.459){\times}10^{981\phantom{0}}$ & $5.16$ & $(1.021 \pm 0.012){\times}10^{982\phantom{0}}$ & $1.21$ & ?\\
IV & $500$ & $1/2$ & $(9.759 \pm 0.650){\times}10^{1131}$ & $6.66$ & $(9.228 \pm 0.228){\times}10^{1131}$ & $2.47$ & ? \\
$A_1$ & $20$ & $1$ & $(9.800 \pm 0.008){\times}10^{32\phantom{00}}$ & $0.08$ & $(9.787 \pm 0.014){\times}10^{32\phantom{00}}$ & $0.14$ & $9.784{\times}10^{32\phantom{00}}$ \\
$A_2$ & $20$ & $1$ & $(3.513 \pm 0.004){\times}10^{32\phantom{00}}$ & $0.10$ & $(3.506 \pm 0.007){\times}10^{32\phantom{00}}$ & $0.21$ & $3.514{\times}10^{32\phantom{00}}$\\
$A_3$ & $15$ & $1/2$ & $(1.456 \pm 0.009){\times}10^{22\phantom{00}}$ & $0.60$ & $(1.437 \pm 0.003){\times}10^{22\phantom{00}}$ & $0.22$ & $1.439{\times}10^{22\phantom{00}}$\\
$A_4$ & $15$ & $1/2$ & $(7.049 \pm 0.044){\times}10^{21\phantom{00}}$ & $0.63$ & $(7.043 \pm 0.022){\times}10^{21\phantom{00}}$ & $0.32$ & $7.034{\times}10^{21\phantom{00}}$\\
$A_5$ & $20$ & $1$ & $(3.290 \pm 0.003){\times}10^{49\phantom{00}}$ & $0.09$ & $(3.294 \pm 0.012){\times}10^{49\phantom{00}}$ & $0.38$ & $3.290{\times}10^{49\phantom{00}}$\\
$A_6$ & $20$ & $1$ & $(5.928 \pm 0.024){\times}10^{40\phantom{00}}$ & $0.40$ & $(5.782 \pm 0.103){\times}10^{40\phantom{00}}$ & $1.72$ & $5.946{\times}10^{40\phantom{00}}$\\
$A_7$ & $15$ & $1/2$ & $(2.069 \pm 0.013){\times}10^{31\phantom{00}}$ & $0.63$ & $(2.092 \pm 0.008){\times}10^{31\phantom{00}}$ & $0.37$ & $2.095{\times}10^{31\phantom{00}}$\\
$A_8$ & $15$ & $1/2$ & $(1.579 \pm 0.020){\times}10^{25\phantom{00}}$ & $1.27$ & $(1.549 \pm 0.027){\times}10^{25\phantom{00}}$ & $1.68$ & $1.579{\times}10^{25\phantom{00}}$\\
$K(x)_9$ & $9$ & $1/2$ & $(4.524 \pm 0.036){\times}10^{0\phantom{000}}$ & $0.79$ & $(4.504 \pm 0.020){\times}10^{0\phantom{000}}$ & $0.43$ & $4.505{\times}10^{0\phantom{000}}$\\ 
$K(x)_{11}$ & $11$ & $1/2$ & $(1.634 \pm 0.014){\times}10^{2\phantom{000}}$ & $0.86$ & $(1.622 \pm 0.009){\times}10^{2\phantom{000}}$ & $0.56$ & $1.623{\times}10^{2\phantom{000}}$\\ 
$K(x)_{13}$ & $13$ & $1/2$ & $(5.815 \pm 0.050){\times}10^{3\phantom{000}}$ & $0.86$ & $(5.844 \pm 0.026){\times}10^{3\phantom{000}}$ & $0.45$ & $5.816{\times}10^{3\phantom{000}}$\\
$K(x)_{15}$ & $15$ & $1/2$ & $(2.134 \pm 0.019){\times}10^{5\phantom{000}}$ & $0.89$ & $(2.117 \pm 0.011){\times}10^{5\phantom{000}}$ & $0.53$ & $2.114{\times}10^{5\phantom{000}}$\\
$K(x)_{100}^{Tr}$ & $100$ & $1/2$ & $(1.876 \pm 0.118){\times}10^{-16\phantom{.}}$ & $6.28$ & $(1.928 \pm 0.037){\times}10^{-16\phantom{.}}$ & $1.90$ & $1.911{\times}10^{-16\phantom{.}}$ 
\end{tabular}}
\label{t:s1}
\end{table}

\subsection{Additional numerical illustrations for the uniform distribution} \label{s:uniform}

Our original interest in these problems was motivated by the uniform distribution over $\Omega^*$ and we have a variety of simulations investigating this special case.  This section is largely reproduced from one of our 2009 unpublished preprints, {\tt ar{X}iv:0906.1004v1}, which focused on comparing different combinatorial approximations and was the basis for our emphasis on the \citet{canfield2008aed} approximation in the main text.  The simulations from this section were carried out in 2009 on a MacBook laptop with 2 GB of RAM and a 2.16 GHz dual core processor using Matlab.  Everything in this section refers to the uniform distribution with $\bs w\equiv 1$.

Supplementary Table \ref{t:time} details the speed of the algorithm on $1000\times 1000$ binary matrices with all row and column sums identical.  These run-times are merely meant to provide a feel for how the algorithm behaves --- no attempt was made to control the other processes operating simultaneously on the laptop.  Presumably a careful C or assembly language implementation would run much faster.  The observed runtime scales closely with the computational complexity of $O(md)$.  So, for example, $100\times 100$ $r_1$-regular matrices can be sampled about $100$ times faster than $1000\times 1000$ $r_1$-regular matrices, and $10\times 10$ matrices can be sampled about $10\mspace{2mu}000$ times faster. 

\begin{table}[h]
\centering
\caption{Sampling time per $1000\times 1000$ $r_1$-regular matrix}
{\small
\begin{tabular}{cccccccccc}
$r_1$ & $2$ & $4$ & $8$ & $16$ & $32$ & $64$ & $128$ & $256$ & $512$ \\ 
time (s) & $1.2$ & $1.6$ & $2.4$ & $4.0$ & $6.7$ & $12.4$ & $24.4$ & $39.2$ & $46.6$ 
\end{tabular}}
\label{t:time}
\end{table}

Supplementary Table \ref{t:cv_1} reports diagnostics on these examples using $T=1000$.  We note that the true number of $1000\times1000$ two-regular matrices is $1.75147\dotsb{\times}10^{5133}$; see Supplementary Section \ref{s:2reg}.  The approximation from the first row of Supplementary Table \ref{t:cv_1} is quite accurate.

\begin{table}[h]
\centering
\caption{Performance for the uniform distribution over $1000\times 1000$ $r_1$-regular binary matrices}
{\small
\begin{tabular}{cccc}
$r_1$ $\vphantom{\bigl(}$ & $\hat\Delta_T$ & $\hat\cv^2_T$ & $\widehat\kappa_T$ \\
$2$  & $0.049$ & $4.2{\times} 10^{-6}$ & $(1.75148 \pm 0.00011) \times 10^{5133\phantom{00}}$ \\ 
$4$  & $0.075$ & $6.4{\times} 10^{-6}$ & $(7.64296 \pm 0.00061) \times 10^{9910\phantom{00}}$ \\ 
$8$  & $0.041$ & $2.1{\times} 10^{-6}$ & $(1.01879 \pm 0.00005) \times 10^{18531\phantom{0}}$ \\ 
$16$  & $0.008$ & $3.9{\times} 10^{-7}$ & $(2.31580 \pm 0.00005) \times 10^{33629\phantom{0}}$ \\ 
$32$  & $0.005$ & $2.3{\times} 10^{-7}$ & $(6.50167 \pm 0.00010) \times 10^{59218\phantom{0}}$ \\ 
$64$ & $0.004$ & $2.2{\times} 10^{-7}$ & $(1.22048 \pm 0.00002) \times 10^{100716}$ \\ 
$128$  & $0.004$ & $1.8{\times} 10^{-7}$ & $(9.38861 \pm 0.00013) \times 10^{163302}$ \\
$256$  & $0.004$ & $2.2{\times} 10^{-7}$ & $(6.70630 \pm 0.00010) \times 10^{243964}$ \\ 
$512$  & $0.004$ & $2.2{\times} 10^{-7}$ & $(5.02208 \pm 0.00007) \times 10^{297711}$
\end{tabular}}
\label{t:cv_1}
\end{table}

\cite{bezakova2006nes} investigates the performance of the \citet{Chen:Sequential:2005} algorithm on pathological margins with very large $r_1$ and $c_1$, but with all other row and column sums exactly $1$.  They prove that the \citet{Chen:Sequential:2005} proposal distribution is extremely far from uniform for such margins, too far for importance sampling to be practical.  It seems likely that our $Q^*$ suffers from the same problem, because of the similarities between the combinatorial approximations in each approach.  The empirical performance of the $Q^*$ from the main text is quite bad in these cases; see below.  On the other hand, it is straightforward to show that using the combinatorial approximations in Supplementary Section \ref{s:c} gives $Q^*=P^*$ for these types of margins.

Following \cite{bezakova2006nes}, we experiment with the margins ${\bs r}^\tr=(240,1,\dotsc,1)$ and ${\bs c}=(179,1,\dotsc,1)$ for a $240\times 301$ matrix.  By conditioning on the entry in the first row and the first column and then using symmetry, one can see that 
\[ N_{m,n}(\bs{r},\bs{c}) = \binom{300}{240}\binom{239}{179}60! + \binom{300}{239}\binom{239}{178}61! = 9.6843\dotsc{\times} 10^{205} . \]
Generating a single observation takes about $0.077$ s. Using $T=10^5$ gives $\hat\Delta_T = 4.1{\times}10^{11}$, $\hat\cv^2_T=1.7{\times}10^3$, and $\hat\kappa_T = (2.2\pm 0.3){\times}10^{205}$, which is quite bad and highly misleading: approximate $95\%$ confidence intervals created by doubling the standard errors would not come close to covering the true value of $\kappa$.  Alternatively, using the algorithm with $u_i$ from Supplementary Section \ref{s:c} gives $\hat\Delta_T=\hat\cv^2=0$ and $\hat\kappa_T = \kappa = 9.6843\dotsb{\times} 10^{205}$, since $Q^*=P^*$ in this case.  In most practical examples, however, the algorithm presented in the main text is superior.

Finally, consider Darwin's finch data \citep[c.f.][]{Chen:Sequential:2005} which is a $13\times 17$ occurrence matrix with $\bs r^\tr=(14, 13, 14, 10, 12,  2, 10, 1, 10, 11,  6,  2, 17)$ and $\bs c=(4,  4, 11, 10,  10, 8,   9,   10, 8, 9$, $3,  10,  4,   7,   9,  3,   3)$.  A single sample takes about $0.001$ s.  With $T=10^6$, we find $\hat\Delta_T=2.8{\times}10^3$ and $\hat\cv^2_T=0.44$ with $\hat\kappa_T=(6.722\pm 0.004){\times}10^{16}$.  \citet{Chen:Sequential:2005} report the true value of $\kappa=  67\mspace{2mu}149\mspace{2mu}106\mspace{2mu}137\mspace{2mu}567\mspace{2mu}626$, and they also report a $\hat\cv^2_T$ of ``around one'' for their algorithm on this problem.  Generally speaking, these importance sampling algorithms tend to be less uniform for small irregular problems like this one, than for the larger and/or more regular examples above.

The previous experiments are based primarily on the internal diagnostics of samples from the proposal distribution $Q^*$.  Other than the asymptotic analysis in \cite{blanchet2006isa} concerning approximate enumeration using a variation of the algorithm of \citet{Chen:Sequential:2005}, there are no external checks on the uniformity of $Q^*$.  Using a complicated, high dimensional proposal distribution without external checks can be dangerous.  Indeed, consider the following worst-case scenario.  Suppose that $\Omega^*=E\cup E^c$, where $E$ is much smaller than $E^c$, and suppose that $Q^*$ is uniform over each of $E$ and $E^c$, but far from uniform over $\Omega^*$, namely,
\[ Q^*(\bs {z}) = \frac{1-\epsilon}{|E|}\ind\{{\bs z}\in E\} + \frac{\epsilon}{|E^c|}\ind\{\bs {z}\in E^c\} \quad  \quad (|E|/|E^c| \ll \epsilon \ll 1) . \]
If $\epsilon$ is extremely tiny, say $\epsilon = 10^{-100}$, then Monte Carlo samples from $Q^*$ will, practically speaking, always lie in $E$, which itself is a tiny fraction of $\Omega^*$.  Furthermore, all internal diagnostics will report that $Q^*$ is exactly uniform, since it is uniform over $E$.  But, of course, statistical inferences based on samples from $Q^*$ will tend to be completely wrong.  This section describes two types of experiments designed to provide external checks on the uniformity of $Q^*$. 

For the first set of experiments we generate a binary matrix $\bs{Z}$ from the uniform distribution over all binary matrices with row sums $\bs{r}$.  This is easy to do by independently and uniformly choosing each row of $\bs{Z}$ from one of the $\binom{n}{r_i}$ possible configurations.  Since the conditional distribution of $\bs{Z}$ given its columns sums $\bs{C}$ is uniform over $\Omega^*_{m,n}(\bs{r},\bs{C})$, we can view $\bs{Z}$ as a single observation from the uniform distribution over $\Omega^*_{m,n}(\bs{r},\bs{C})$.  Of course, there is no practical way to uniformly and independently generate another such $\bs{Z}$ with the same $\bs{C}$.  Notice that the importance weight $f(\bs Z)$ gives external information about the uniformity of $Q^*$ for these margins, since it gives the value of $1/Q^*$ at a uniformly chosen location in $\Omega^*_{m,n}(\bs{r},\bs{C})$.  Indeed, in the pathological thought experiment described above, $\bs{Z}$ would almost certainly be in $E^c$ and $f(\bs Z)$ would be substantially larger than any of the importance weights.  Alternatively, if $Q^*$ is nearly uniform, then $f(\bs Z)$ should be indistinguishable from the other importance weights.  In summary, we can compare $Q^*$ to $P^*$ by comparing the importance weights to $f(\bs Z)$.  This observation can also be used to give valid Monte Carlo p-values with importance sampling, even if the importance sampling distribution is far from the target distribution \citep{harrison2012conservative}.

Each experiment of this type proceeds identically.  We fix $m$, $n$, and $\bs{r}$.  Then we generate $L$ iid observations, say ${\bs Z_0^{(1)}},\dotsc,{\bs Z_0^{(L)}}$, from the uniform distribution over all $m\times n$ binary matrices with row sums $\bs{r}$.  The column sums of these matrices are ${\bs C^{(1)}},\dotsc,{\bs C^{(L)}}$.  Then, for each $\ell=1,\dotsc,L$, we generate $T$ iid observations, say ${\bs Z^{(\ell)}_1},\dotsc,{\bs Z^{(\ell)}_T}$, from the proposal distribution $Q^*$ over $\Omega^*_{m,n}(\bs{r},{\bs C^{(\ell)}})$.  We compute the ratio of maximum to minimum importance weights {\em including the original observation} for each $\ell$, namely, 
\[ \hat\Delta_T^{(\ell)} = \frac{\max_{t=0,\dotsc,T} f(\bs Z^{(\ell)}_t)}{\min_{t=0,\dotsc,T} f(\bs Z^{(\ell)}_t)} - 1 ,\]
and we report the final summary ${\hat\Delta_T}^{\max}  = \max_{\ell=1,\dots,L} \hat\Delta_T^{(\ell)}$.  If $\hat\Delta_T^{\max}$ is close to zero, then this provides evidence that $Q^*$ is approximately uniform over a large part of each $\Omega^*_{m,n}(\bs{r},{\bs C^{(\ell)}})$. 

We begin with $1000\times 1000$ matrices with regular row sums $r_1=\dotsb=r_m$, but the column sums will not be regular, since they are generated randomly.  We use $L=10$ and $T=10$ for the cases $r_1=2,8,32$, finding $\hat\Delta_T^{\max}=0.0002, 0.0023, 0.0051$, respectively.
For another example, take the row sums for the irregular $50\times 100$ case that was used for Table 2 in the main text and take $k=1$, i.e., $\bs{r}={\tilde{\bs r}}$.   We use $L=100$ and $T=1000$ and find that $\hat\Delta_T^{\max}= 1.264$.  These preliminary experiments are encouraging, and suggest that $Q^*$ is indeed a good approximation of uniform $P^*$ in many cases.    
 
 For the second type of experiment, we try to design an extreme $\bs z'\in\Omega^*$ and compare the importance weights to $f(\bs z')$.  Again, if $Q^*$ is approximately uniform over all of $\Omega^*$ then $f(\bs z')$ should be indistinguishable from the other importance weights.  For these experiments we report
\[ {\hat\Delta_T}' = \frac{\max\left\{f(\bs z'),f(\bs Z_1),\dotsc,f(\bs Z_T)\right\}}{\min\left\{f(\bs z'
),f(\bs Z_1),\dotsc,f(\bs Z_T)\right\}} - 1 , \]
which should be close to zero if the region in $\Omega^*$ where $Q^*$ is approximately uniform includes $\bs z'$.

Consider the regular case where $m=n=1000$ and $r_1=r_i=c_j$ for all $i,j$.  Suppose that $r_1$ evenly divides $1000$ and let $\bs z'$ be comprised only of disjoint $r_1\times r_1$ blocks of ones.  In particular, take $\bs z'_{ij}=1$ for $(k-1)r_1+1 \leq i,j \leq kr_1$ and for $k=1,\dotsc,1000/r_1$.  For the cases $r_1=2,4,8$ we compute $f(\bs z')$ and compare it to the data that generated the corresponding parts of table \ref{t:cv_1}, obtaining $\hat\Delta_T'=0.741, 26.24, 6.25{\times} 10^4$, respectively.  Clearly, $Q^*$ is not a uniformly accurate approximation of $P^*$ over all of $\Omega^*$ and is unlikely to be useful as a proposal distribution for rejection sampling to get exact samples from $P^*$.  Nevertheless, $Q^*$ seems to be extremely well-suited as a proposal distribution for importance sampling.  For another example, consider the irregular $50\times 100$ case that was used for Table 2 in the main text and take $k=1$, i.e., ${\bs r}={\tilde{\bs  r}}$ and $\bs c=\tilde {\bs c}$.  We construct a pathological ${\bs z'}$ as follows.  Place $c_1$ ones in the last $c_1$ rows, corresponding to the smallest row sums, of the first column.  Place $c_2$ ones in the last available $c_2$ rows of the second column, where a row is available if placing a one in that row will not exceed the row sum for that row.  Continue in this manner until all the columns are assigned or until a column cannot be assigned successfully.  In general, this procedure is not guaranteed to terminate successfully, but it does for this choice of margins.  The resulting ${\bs z'}$ is unusual because rows and columns with large sums tend to have zeros at the intersecting entry.  Using the data from the corresponding part of Table 2 in the main text gives ${\hat\Delta_T}' = 14.37$.


\section*{Supplementary Appendix}

\small

\subsection*{The number of $100\times 100$ two-regular binary matrices}

\begin{center}
\noindent \begin{tabular}{l@{ }l@{ }l@{ }l@{ }l@{ }l@{ }l@{ }l}
2969298425&4879211020&5463258948&9046531125&6932010720&0899043082&6661472985&5602957737\\5386603250&7914169840&3947972542&0803105057&9494091210&8196163985&3132939771&8223074880\\1582489734&4113002630&0345104451&5505567811&8301236764&6670284335&5753266570&2919415207\\2361422613&1731302283&4023510256&2089359423&4174989926&4000000000&0000000000&00000
\end{tabular}
\end{center}

\subsection*{The number of $500\times 500$ two-regular binary matrices}

\begin{center}
\noindent \begin{tabular}{l@{ }l@{ }l@{ }l@{ }l@{ }l@{ }l@{ }l}
2276586004&3872645654&7163822917&6140246378&6529219189&6007058852&1885701633&9308224336\\7024859918&5873168947&8428993358&7710991052&6831024823&1020957186&1359882527&3634597638\\7751901014&9459428637&5300752209&6236400145&2272455600&2450498447&6886449802&2657577100\\8803085437&1426603063&9060350752&5676829379&2441654640&4384402364&9178512515&5701834312\\5382285704&7911170936&9213162976&1124369611&0263144354&2492660647&6317501009&4702298551\\3783877264&5366936440&0850289755&0247749665&4582496735&4778933695&9359401807&4728987947\\4052084791&8351006525&6516882276&6819426986&4276522770&8754690714&8153703130&7689579335\\5313886817&9879619523&6757312609&9563935644&5860973860&5720751902&8525628015&1655464790\\3607836217&2202522127&9381851238&5339132917&8663772909&4697618230&9562268584&1389355037\\4200343275&4426328049&4429348983&4734923700&0635594018&1200043308&9996436581&2082429967\\1420144526&3238392163&0625410465&1147246306&0267066287&2838455102&1984436331&4795820153\\4878729606&4682614593&4828351763&2549610945&2823414530&6966187549&3636469942&1582542169\\0511243887&9654470644&8952801709&4100687806&1803581920&0502635810&6084543151&8196763100\\9226192052&8186323173&8128828855&7307283447&5486503911&0996089630&7969624574&8668199425\\1430690842&9240854111&3288457886&5068062328&1130147009&2410850737&0194640624&5215023611\\0105313331&5631006370&7547904555&8541951209&3762970404&4299114208&6898539174&1261578007\\5271576323&7806458898&5197173413&2333790169&8450503603&6175432120&4646913329&9283772618\\0789892314&7885014128&9831206980&1470933069&2885920165&3886059912&3547627990&2473766270\\0084914243&1261925800&3966112818&5515090740&2869173796&5265773700&6653705150&0776999823\\6682749949&6649629337&6729065663&7740220752&0069908832&1026134189&8109544591&4141299020\\9944691129&8101632276&5735759559&3131678694&4342947732&7389063830&1146871076&6098180223\\5086650691&0193318778&3650834389&5788540935&3233656140&3425148468&8948999361&5539721393\\2767810044&6245991329&5809908199&9005968612&6446584189&0334076925&7082772956&3377889631\\0446650398&8183375905&5124117054&7434261832&8900372657&5745038153&2952534928&4112016395\\9467531245&7165626500&2517876951&1088955612&4288697963&9375087520&6487400471&4382991165\\8206541306&8546637026&9648941941&8803223917&8589969888&6361729999&1147924387&2385375087\\0828596942&2197021633&2700563010&0820849326&1167561772&1388697124&8640000000&0000000000\\0000000000&0000000000&0000000000&0000000000&0000000000&0000000000&0000000000&0000000000\\0000000000&0000000000&0000000
\end{tabular}
\end{center}

\subsection*{Exactly computing the $\alpha$-permanent of a constant matrix}

If $\bs \pi=(\pi_1,\dotsc,\pi_n)$ is a permutation chosen uniformly at random and $C$ is the number of disjoint cycles in $\bs \pi$, then $C$ has the same distribution as $B_1+\dotsb+B_n$, where each $B_i$ is independent Bernoulli$(1/i)$ \citep[Lemma 2.2.5]{durrett2010probability}.  If $\bs w$ is an $n\times n$ constant matrix with common entry $b$, then
\[ \text{per}_\alpha(\bs w) =  n!b^n\Exp(\alpha^C) = n!b^n\prod_{i=1}^n\Exp(\alpha^{B_i}) = n!b^n \prod_{i=1}^n \Bigl(\frac{\alpha}{i}+\bigl(1-\frac{1}{i}\bigr)\Bigr) = n!b^n \prod_{i=1}^n \frac{i + \alpha-1}{i} . \]
We used this formula with $n=500$ and $b=1$ to get the true value of $\text{per}_\alpha(\bs w)$ for $\bs w$ in class I in Supplementary Table \ref{t:s1}.

\subsection*{Matlab implementation of the algorithm}

This is a place-holder for cleaner, shorter code that will be inserted prior to publication.  Software will also be available on the author's website.

\tiny

\begin{verbatim}
function [logQ,logP,alist] = BernoulliMarginsRnd(SampN,rN,cN,wN,pflag,wflag,cflag,bIN)
%function [logQ,logP,alist] = BernoulliMarginsRnd(N,r,c,w,pflag,wflag,cflag,Binput)
%
% Approximate sampling from independent Bernoulli random variables B(i,j)
% arranged as an m x n matrix B given the m-vector of row sums r and the
% n-vector of column sums c, i.e., given that sum(B,2)=r and sum(B,1)=c.
%
% An error is generated if no binary matrix agrees with r and c.
%
% B(i,j) is Bernoulli(p(i,j)) where p(i,j)=w(i,j)/(1+w(i,j)), i.e.,
% w(i,j)=p(i,j)/(1-p(i,j)).  [The case p(i,j)=1 must be handled by the user
% in a preprocessing step, by converting to p(i,j)=0 and decrementing the
% row and column sums appropriately.]  
%
% Use w=[] for w identically 1, i.e., approximate uniform sampling over
% binary matrices with margins r and c.
%
% N is the sample size.  Because of pre-processing, it is more efficient
% per matrix to use larger sample sizes.
% 
% alist stores the locations of the ones in the samples.  
% If d = sum(r) = sum(c), then alist is 2 x d x N.
%
% The 1-entries of the kth matrix are stored as alist(:,:,k).  The
% (row,column) indices are (alist(1,t,k),alist(2,t,k)) for t=1:d.
% 
% If B is the kth matrix, then B can be created from alist via:
%
% B = false(m,n); for t = 1:size(alist,2), B(alist(1,t,k),alist(2,t,k)) = true; end
%
% logQ(k)=log(probability that algorithm generates B)
% logP(k)=log(prod(w(B)))
%
% If the algorithm is used for importance sampling, then the kth
% unnormalized importance weight is exp(logP(k)-logQ(k)).
%
% NOTE for w(i,j)=0:
%
% If the entries of w are not strictly positive, then the algorithm can 
% sometimes generate matrices with logP(k)=-inf.  In these cases, some of
% the entries of alist(:,:,k) may be zero and logQ(k) corresponds to the
% probability of generating that particular alist(:,:,k).
%
% OPTIONS:
%
% pflag: 'canfield' or '' (default, works best in most cases)
%        'greenhill' (perhaps useful for sparse and highly irregular margins)
% pflag controls which combinatorial approximations are used
%
% wflag: 'sinkhorn' or '' (default)
% wflag controls the initial balancing of w; it is passed to canonical.m
%
% cflag: 'descend' or '' (default)
%        'none' (sample columns in original order)
% cflag controls the order in which the columns are sampled
%
% Binput is a m x n binary matrix.  If it is provided, then the algorithm
% computes the probability of generating this matrix.

if nargin < 8 || isempty(bIN)
    doIN = false;
else
    doIN = true;
end
if nargin < 7 || isempty(cflag)
    cflag = 'descend';
end
if nargin < 6 || isempty(wflag)
    wflag = 'sinkhorn';
end
if nargin < 5 || isempty(pflag)
    pflag = 'canfield';
end
if nargin < 4
    wN = [];
end

doW = true;
if isempty(wN), doW = false; end

doA = true;
if nargout < 2, doA = false; end

if ~isscalar(SampN) || SampN < 1 || SampN ~= round(SampN), error('SampN must be a positive integer'), end

ptype = 0;
switch lower(pflag)
    case 'canfield'
        ptype = 1;
    case 'greenhill'
        ptype = 2;
    otherwise
        error('unknown pflag')
end

%------------------------------------------------------%
%--------------- START: PREPROCESSING -----------------%
%------------------------------------------------------%

% sizing
mT = numel(rN);
nT = numel(cN);

% sort the marginals (descending)
rT = rN(:);
[rsort,rndxT] = sort(rT,'descend');

if doW
    % balance the weights
    [~,~,wopt] = canonical(wN,wflag);
    % reorder the columns
    switch lower(cflag)
        case 'none'
            cndx = 1:nT;
        case 'descend'
            [~,cndx] = sortrows(-[cN(:) var(wopt,0,1).']);
        otherwise
            error('unknown cflag')
    end
    csort = cN(cndx);
    wopt = wopt(:,cndx);
    % precompute log weights
    logw = log(wN);

    % ----------------------------------------------------
    % precompute G
    
    logwopt = log(wopt);
    
    rmax = max(rT);
    G = -inf(rmax+1,mT,nT-1);
    G(1,:,:) = 0;
    G(2,:,nT-1) = logwopt(:,nT);
    
    for i = 1:mT
        ri = rT(i);
        for j = nT-1:-1:2
            wij = logwopt(i,j);
            for k = 2:ri+1
                b = G(k-1,i,j)+wij;
                a = G(k,i,j);
                if a > -inf || b > -inf
                    if a > b
                        G(k,i,j-1) = a + log(1+exp(b-a));
                    else
                        G(k,i,j-1) = b + log(1+exp(a-b));
                    end
                end
            end
        end
        
        for j = 1:nT-1
            for k = 1:rmax
                Gknum = G(k,i,j);
                Gkden = G(k+1,i,j);
                if isinf(Gkden)
                    G(k,i,j) = -1;
                else
                    G(k,i,j) = wopt(i,j)*exp(Gknum-Gkden)*((nT-j-k+1)/k);
                end
            end
            if isinf(Gkden)
                G(rmax+1,i,j) = -1;
            end
        end
    end
    % ----------------------------------------------------
else
    switch lower(cflag)
        case 'none'
            cndx = 1:numel(cN);
        case 'descend'
            [csort,cndx] = sort(cN(:),'descend');
        otherwise
            error('unknown cflag')
    end
end

% generate the inverse index for the row orders to facilitate fast
% sorting during the updating
irndxT = (1:mT).'; irndxT(rndxT) = irndxT;

% basic input checking
if rsort(1) > nT || rsort(mT) < 0 || csort(1) > mT || csort(nT) < 0 || any(rsort ~= round(rsort)) || any(csort ~= round(csort))
   error('marginal entries invalid')
end

% compute the conjugate of c
cconjT = conjugate_local(csort,mT);

% get the running total of number of ones to assign
countT = sum(rsort);

% get the running total of sum of c squared
ccount2T = sum(csort.^2);
% get the running total of (2 times the) column marginals choose 2
ccount2cT = sum(csort.*(csort-1));
% get the running total of (6 times the) column marginals choose 3
ccount3cT = sum(csort.*(csort-1).*(csort-2));

% get the running total of sum of r squared
rcount2T = sum(rsort.^2);
% get the running total of (2 times the) row marginals choose 2
rcount2cT = sum(rsort.*(rsort-1));
% get the running total of (6 times the) row marginals choose 3
rcount3cT = sum(rsort.*(rsort-1).*(rsort-2));

% check for compatible marginals
if countT ~= sum(csort) || any(cumsum(rsort) > cumsum(cconjT)), error('marginal sums invalid'), end

% initialize the memory
logQ = zeros(SampN,1);
logP = zeros(SampN,1);
if doA, AN = SampN; else AN = 1; end
alist = zeros(2,countT,AN);
% initialize the memory
M = csort(1)+3; % index 1 corresponds to -1; index 2 corresponds to 0, index 3 corresponds to 1, ..., index M corresponds to c(1)+1
S = zeros(M,nT);
SS = zeros(M,1);

eps0 = eps(0); % used to prevent divide by zero

%------------------------------------------------------%
%--------------- END: PREPROCESSING -------------------%
%------------------------------------------------------%

% loop over the number of samples
for SampLoop = 1:SampN
 
    %--------------- INITIALIZATION -----------------------%
    if doA, ALoop = SampLoop; else ALoop = 1; end
    
    % copy in initialization
    r = rT;
    rndx = rndxT;
    irndx = irndxT;
    
    cconj = cconjT;
    count = countT;
    ccount2 = ccount2T;
    ccount2c = ccount2cT;
    ccount3c = ccount3cT;
    rcount2 = rcount2T;
    rcount2c = rcount2cT;
    rcount3c = rcount3cT;
    m = mT;
    n = nT;
    
    % initialize
    place = 0; % most recent assigned column in alist
    logq = 0; % running log probability
    logp = 0;
    
    %------------------------------------------------------%
    %--------------- START: COLUMN-WISE SAMPLING ----------%
    %------------------------------------------------------%
    
    %-------- loop over columns ------------%
    for c1 = 1:nT
                
        %-----------------------------------------------------------------%
        %------------- START: SAMPLE THE NEXT "COLUMN" -------------------%
        %-----------------------------------------------------------------%
        
        % remember the starting point for this columns
        placestart = place + 1;
        
        %--------------------------------
        % sample a col
        %--------------------------------
        
        label = cndx(c1); % current column label
        
        colval = csort(c1); % current column value
        
        if colval == 0 || count == 0, break, end
        
        % update the conjugate
        for i = 1:colval
            cconj(i) = cconj(i)-1;
        end
        % update the number of columns remaining
        n = n - 1;
        
        %------------ DP initialization -----------
        
        smin = colval;
        smax = colval;
        cumsums = count;
        % update the count
        count = count - colval;
        % update running total of sum of c squared
        ccount2 = ccount2 - colval^2;
        % update the remaining (two times the) sum of column sums choose 2
        ccount2c = ccount2c - colval*(colval-1);
        % update the remaining (six times the) sum of column sums choose 3
        ccount3c = ccount3c - colval*(colval-1)*(colval-2);
        
        cumconj = count;
        
        SS(colval+3) = 0;
        SS(colval+2) = 1;
        SS(colval+1) = 0;
        
        % get the constants for computing the probabilities
        % it is faster to compute them all, than to check pflag
        d = ccount2c/count^2;
        if (count == 0) || (m*n == count)
            weightA = 0;
        else
            weightA = m*n/(count*(m*n-count));
            weightA = weightA*(1-weightA*(ccount2-count^2/n))/2;
        end
        
        d2 = ccount2c/(2*count^2+eps0) + ccount2c/(2*count^3+eps0) + ccount2c^2/(4*count^4+eps0);
        d3 = -ccount3c/(3*count^3+eps0) + ccount2c^2/(2*count^4+eps0);
        d22 = ccount2c/(4*count^4+eps0) + ccount3c/(2*count^4+eps0) - ccount2c^2/(2*count^5+eps0);
        
        %----------- dynamic programming ----------
        SSS = 0;
        % loop over (remaining and sorted descending) rows in reverse
        for i = m:-1:1
            
            % get the value of this row and use it to compute the
            % probability of a 1 for this row/column pair
            rlabel = rndx(i);
            val = r(rlabel);
            if ptype == 1
                % canfield
                p = val*exp(weightA*(1-2*(val-count/m)));
                p = p./(n+1-val+p);
                q = 1-p;
            elseif ptype == 2
                % greenhill
                q = 1/(1+val*exp((2*d2+3*d3*(val-2)+4*d22*(rcount2c-val+1))*(val-1)));
                p = 1-q;
            else
                % never get here
                p = 0; q = 0; % helps compiler
            end
            
            % incorporate weights
            if doW && n > 0 && val > 0
                Gk = G(val,rlabel,c1);
                if Gk < 0
                    q = 0;
                else
                    p = p*Gk;  
                end
            end
            
            % update the feasibility constraints
            cumsums = cumsums - val;
            cumconj = cumconj - cconj(i);
            
            sminold = smin;
            smaxold = smax;
            
            % incorporate the feasibility constraints into bounds on the
            % running column sum
            smin = max(0,max(cumsums-cumconj,sminold-1));
            smax = min(smaxold,i-1);
            
            % DP iteration
            SSS = 0;
            
            SS(smin+1) = 0;  % no need to set S(1:smin) = 0, since it is not accessed
            for j = smin+2:smax+2
                a = SS(j)*q;
                b = SS(j+1)*p;
                apb = a + b;
                SSS = SSS + apb;
                SS(j) = apb;
                S(j,i) = b/(apb+eps0);
            end
            SS(smax+3) = 0;  % no need to set S(smax+4:end) = 0, since it is not accessed
            
            % check for impossible
            if SSS <= 0, break, end
            
            % normalize to prevent overflow/underflow
            for j = smin+2:smax+2
                SS(j) = SS(j) / SSS;
            end
            
        end
        
        % check for impossible
        if SSS <= 0, logp = -inf; break, end
        
        %----------- sampling ----------
        j = 2; % running total (offset to match indexing offset)
        jmax = colval + 2;
        if j < jmax % skip assigning anything when colval == 0
            if doIN
                for i = 1:m
                    % get the transition probability of generating a one
                    p = S(j,i);
                    % get the current row
                    rlabel = rndx(i);
                    if bIN(rlabel,label)
                        
                        % if we have a generated a one, then decrement the current
                        % row total
                        val = r(rlabel);
                        r(rlabel) = val-1;
                        
                        rcount2 = rcount2 - 2*val + 1;
                        rcount2c = rcount2c - 2*val + 2;
                        rcount3c = rcount3c - 3*(val-1)*(val-2);
                        
                        % record the entry and update the log probability
                        place = place + 1;
                        logq = logq + log(p);
                        if doW, logp = logp + logw(rlabel,label); end
                        alist(1,place,ALoop) = rlabel;
                        alist(2,place,ALoop) = label;
                        j = j + 1;
                        % the next test is not necessary, but seems more efficient
                        % since all the remaining p's must be 0
                        if j == jmax, break, end
                    else
                        logq = logq + log(1-p);
                    end
                end
            else
                for i = 1:m
                    % get the transition probability of generating a one
                    p = S(j,i);
                    if rand <= p
                        
                        % if we have a generated a one, then decrement the current row total
                        rlabel = rndx(i);
                        val = r(rlabel);
                        r(rlabel) = val-1;
                        
                        rcount2 = rcount2 - 2*val + 1;
                        rcount2c = rcount2c - 2*val + 2;
                        rcount3c = rcount3c - 3*(val-1)*(val-2);
                        
                        % record the entry and update the log probability
                        place = place + 1;
                        logq = logq + log(p);
                        if doW, logp = logp + logw(rlabel,label); end
                        alist(1,place,ALoop) = rlabel;
                        alist(2,place,ALoop) = label;
                        j = j + 1;
                        % the next test is not necessary, but seems more efficient
                        % since all the remaining p's must be 0
                        if j == jmax, break, end
                    else
                        logq = logq + log(1-p);
                    end
                end
            end
        end
        
        %-----------------------------------------------------------------%
        %------------- END: SAMPLE THE NEXT "COLUMN" ---------------------%
        %-----------------------------------------------------------------%
        
        if count == 0, break, end
        
        %-----------------------------------------------
        % everything is updated except the sorting
        %-----------------------------------------------
        
        %-----------------------------------------------------------------%
        %------------- START: RESORT THE NEW ROW SUMS --------------------%
        %-----------------------------------------------------------------%
        
        % re-sort the assigned rows
        
        % this code block takes each row that was assigned to the list
        % and either leaves it in place or swaps it with the last row
        % that matches its value; this leaves the rows sorted (descending)
        % since each row was decremented by only 1
        
        % looping in reverse ensures that least rows are swapped first
        for j = place:-1:placestart
            % get the row label and its new value (old value -1)
            k = alist(1,j,ALoop);
            val = r(k);
            % find its entry in the sorting index
            irndxk = irndx(k);
            % look to see if the list is still sorted
            irndxk1 = irndxk + 1;
            if irndxk1 > m || r(rndx(irndxk1)) <= val
                % no need to re-sort
                continue;
            end
            % find the first place where k can be inserted
            irndxk1 = irndxk1 + 1;
            while irndxk1 <= m && r(rndx(irndxk1)) > val
                irndxk1 = irndxk1 + 1;
            end
            irndxk1 = irndxk1 - 1;
            % now swap irndxk and irndxk1
            rndxk1 = rndx(irndxk1);
            rndx(irndxk) = rndxk1;
            rndx(irndxk1) = k;
            irndx(k) = irndxk1;
            irndx(rndxk1) = irndxk;
        end
        
        %-----------------------------------------------------------------%
        %------------- END: RESORT THE NEW ROW SUMS ----------------------%
        %-----------------------------------------------------------------%
        
        % r(rndx(rndx1:rndxm)) is sorted descending and has exactly those
        % unassigned rows
        % rndx(rndx1:rndxm) still gives the labels of those rows
        % rndx(irndx(k)) = k
        %
        % c(c1+1:cn) is sorted descending and has exactly those unassigned columns
        % cndx(c1+1:cn) still gives the labels of those columns
        %
        % m, n, count, ccount2, ccount2c are valid for the remaining rows, cols
        
    end
    
    logQ(SampLoop) = logq;
    logP(SampLoop) = logp;
    
end

%-------------------------------------------------------------------------%
%-------------------------------------------------------------------------%
%-------------------------------------------------------------------------%
%------------------ END OF MAIN FUNCTION ---------------------------------%
%-------------------------------------------------------------------------%
%-------------------------------------------------------------------------%
%-------------------------------------------------------------------------%

% helper function (just to keep everything together... not for efficiency,
% since it is only called once)

function cc = conjugate_local(c,n)
% function cc = conjugate(c,n)
%
% let c(:) be nonnegative integers
% cc(k) = sum(c >== k)  for k = 1:n

cc = zeros(n,1);

%c = min(c,n);

for j = 1:numel(c)
    k = c(j);
    if k >= n
        cc(n) = cc(n) + 1;
    elseif k >= 1
        cc(k) = cc(k) + 1;
    end
end

s = cc(n);
for j = n-1:-1:1
    s = s + cc(j);
    cc(j) = s;
end

%-----------------------------------

function [a,b,abw,k] = canonical(w,flag,tol,maxiter,r,c)

[m,n] = size(w);

if nargin <6 || isempty(c)
	c = ones(1,n);
elseif size(c,1) ~= 1
	c = c(:).';
end
if nargin <5 || isempty(r)
	r = ones(m,1);
elseif size(r,2) ~= 1
	r = r(:);
end
if nargin <4 || isempty(maxiter)
    maxiter = 10^5;
end
if nargin <3 || isempty(tol)
    tol = 1e-8;
end
if nargin <2 || isempty(flag)
    flag = 'sinkhorn';
end

switch lower(flag)
    
    case 'sinkhorn'
              
		M = sum(w>0,1); N = sum(w>0,2);
		a = N./sum(w,2); a = a/mean(a);		
		b = M./sum(bsxfun(@times,a,w),1);
		
		if tol >= 0, a0 = a; b0 = b; end

        k = 0;
        tolcheck = inf;
        while k < maxiter && tolcheck > tol
            k = k + 1;
		
			a = N./sum(bsxfun(@times,b,w),2); a = a/mean(a);
			b = M./sum(bsxfun(@times,a,w),1);
			
			if tol >= 0
                tolcheck = sum(abs(a-a0))+sum(abs(b-b0));
				a0 = a; b0 = b;
			end
        end 
             
    case 'sinkhorn-col'
              
		w = fliplr(w);
		
		M = sum(w>0,1); N = cumsum(w>0,2);
		aa = N./cumsum(w,2);
		b = M./sum(w.*aa,1); b = b / mean(b);
		a = aa(:,n);
				
		if tol >= 0, a0 = a; b0 = b; end

        k = 0;
        tolcheck = inf;
        while k < maxiter && tolcheck > tol
            k = k + 1;
		
			aa = N./cumsum(bsxfun(@times,b,w),2);
			b = M./sum(w.*aa,1); b / mean(b);
			a = aa(:,n);
			
			if tol >= 0
                tolcheck = sum(abs(a-a0))+sum(abs(b-b0));
				a0 = a; b0 = b;
			end
		end
		
		w = fliplr(w);
		b = fliplr(b);
		
	case 'log'
		
		w0 = w > 0;
		M = sum(w0,1); N = sum(w0,2);
		logw = log(w+~w0);
		a = exp(-sum(logw,2)./N);
		b = exp(-sum(logw,1)./M);
		
	case 'entropy'
		
		w1 = (w > 0)./max(w,eps(0));
		a = sqrt(sum(w1,2)./sum(w,2)); a = a/mean(a);
		b = sqrt(sum(bsxfun(@rdivide,w1,a),1)./sum(bsxfun(@times,a,w),1));
		
		if tol >= 0, a0 = a; b0 = b; end

        k = 0;
        tolcheck = inf;
        while k < maxiter && tolcheck > tol
            k = k + 1;
		
			a = sqrt(sum(bsxfun(@rdivide,w1,b),2)./sum(bsxfun(@times,b,w),2)); a = a/mean(a);
			b = sqrt(sum(bsxfun(@rdivide,w1,a),1)./sum(bsxfun(@times,a,w),1));
			
			if tol >= 0
                tolcheck = sum(abs(a-a0))+sum(abs(b-b0));
				a0 = a; b0 = b;
			end
        end 
		
	case 'l2'
              
		w2 = w.^2;
		
		a = sum(w,2)./sum(w2,2); a = a/mean(a);		
		b = sum(bsxfun(@times,a,w),1)./sum(bsxfun(@times,a.^2,w2),1);
		
		if tol >= 0, a0 = a; b0 = b; end

        k = 0;
        tolcheck = inf;
        while k < maxiter && tolcheck > tol
            k = k + 1;
		
			a = sum(bsxfun(@times,b,w),2)./sum(bsxfun(@times,b.^2,w2),2); a = a/mean(a);
			b = sum(bsxfun(@times,a,w),1)./sum(bsxfun(@times,a.^2,w2),1);
			
			if tol >= 0
                tolcheck = sum(abs(a-a0))+sum(abs(b-b0));
				a0 = a; b0 = b;
			end
        end 
             	
    case 'l2p'
        
        w2 = w.^2;
        
        c = (1+w).^3;
		a = sum(w./c,2)./sum(w2./c,2); a = a/mean(a);		
        c = (1+bsxfun(@times,a,w)).^3;
		b = sum(bsxfun(@times,a,w)./c,1)./sum(bsxfun(@times,a.^2,w2)./c,1);
		
		if tol >= 0, a0 = a; b0 = b; end

        k = 0;
        tolcheck = inf;
        while k < maxiter && tolcheck > tol
            k = k + 1;
		
            c = (1+a*b.*w).^3;
			a = sum(bsxfun(@times,b,w)./c,2)./sum(bsxfun(@times,b.^2,w2)./c,2); a = a/mean(a);
            c = (1+a*b.*w).^3;
			b = sum(bsxfun(@times,a,w)./c,1)./sum(bsxfun(@times,a.^2,w2)./c,1);
			
			if tol >= 0
                tolcheck = sum(abs(a-a0))+sum(abs(b-b0));
				a0 = a; b0 = b;
			end
        end 
        
    case 'ratio'
        
        wz = w > 0;
        w(~wz) = eps(0);
        
        a = sqrt(sum(wz./w,2)./sum(w,2)); a = a/mean(a);
		b = sqrt(sum(wz./(bsxfun(@times,a,w)),1)./sum(bsxfun(@times,a,w),1));
		
		if tol >= 0, a0 = a; b0 = b; end

        k = 0;
        tolcheck = inf;
        while k < maxiter && tolcheck > tol
            k = k + 1;
		
			a = sqrt(sum(wz./(bsxfun(@times,b,w)),2)./sum(bsxfun(@times,b,w),2)); a = a/mean(a);
            b = sqrt(sum(wz./(bsxfun(@times,a,w)),1)./sum(bsxfun(@times,a,w),1));
			
			if tol >= 0
                tolcheck = sum(abs(a-a0))+sum(abs(b-b0));
				a0 = a; b0 = b;
			end
        end 
        
	case 'barvinok'

		s = log(r/n);
		t = log(c/m);

		M = w.*(exp(s)*exp(t));
		M = M ./ (1+M);

		sMr = sum(M,2)-r;
		sMc = sum(M,1)-c;

		tolcheck = sum(abs(sMr))+sum(abs(sMc));

		alpha = .01;
		
		while tolcheck > tol
    
			s = s - alpha*sMr;
			t = t - alpha*sMc;
    
			M = w.*(exp(s)*exp(t));
			M = M ./ (1+M);
    
			sMr = sum(M,2)-r;
			sMc = sum(M,1)-c;
    
			tolcheck = sum(abs(sMr))+sum(abs(sMc));
		end
		
		a = exp(s);
		b = exp(t);
		
    otherwise
        
        error('unknown flag')
end

if nargout > 2, abw = a*b.*w; end
\end{verbatim}



\bibliographystyle{apacite}
\bibliography{BinaryMatrix}	

\begin{thebibliography}{}

\bibitem[\protect\citeauthoryear{%
Admiraal%
\ \BBA{} Handcock%
}{%
Admiraal%
\ \BBA{} Handcock%
}{%
{\protect\APACyear{2008}}%
}]{%
admiraal2008networksis}%
\APACinsertmetastar{%
admiraal2008networksis}%
Admiraal, R.%
\BCBT{}\ \BBA{} Handcock, M\BPBI S.%
%
\unskip\
\newblock
\APACrefYearMonthDay{2008}{}{}.
\newblock
\BBOQ{}\APACrefatitle{networksis: a package to simulate bipartite graphs with
  fixed marginals through sequential importance sampling}{networksis: a package
  to simulate bipartite graphs with fixed marginals through sequential
  importance sampling}.\BBCQ{}
\newblock
\APACjournalVolNumPages{J. Statist. Software}{24}{8}{1--21}.
\PrintBackRefs{\CurrentBib}

\bibitem[\protect\citeauthoryear{%
Anand%
, Dumir%
\BCBL{}\ \BBA{} Gupta%
}{%
Anand%
\ \protect\BOthers{.}}{%
{\protect\APACyear{1966}}%
}]{%
anand1966combinatorial}%
\APACinsertmetastar{%
anand1966combinatorial}%
Anand, H.%
, Dumir, V\BPBI C.%
\BCBL{}\ \BBA{} Gupta, H.%
%
\unskip\
\newblock
\APACrefYearMonthDay{1966}{}{}.
\newblock
\BBOQ{}\APACrefatitle{A combinatorial distribution problem}{A combinatorial
  distribution problem}.\BBCQ{}
\newblock
\APACjournalVolNumPages{Duke Math. J.}{33}{4}{757--769}.
\PrintBackRefs{\CurrentBib}

\bibitem[\protect\citeauthoryear{%
Ando%
}{%
Ando%
}{%
{\protect\APACyear{1989}}%
}]{%
ando1989majorization}%
\APACinsertmetastar{%
ando1989majorization}%
Ando, T.%
%
\unskip\
\newblock
\APACrefYearMonthDay{1989}{}{}.
\newblock
\BBOQ{}\APACrefatitle{Majorization, doubly stochastic matrices, and comparison
  of eigenvalues}{Majorization, doubly stochastic matrices, and comparison of
  eigenvalues}.\BBCQ{}
\newblock
\APACjournalVolNumPages{Linear Algebra Appl.}{118}{}{163--248}.
\PrintBackRefs{\CurrentBib}

\bibitem[\protect\citeauthoryear{%
Bapat%
\ \BBA{} Beg%
}{%
Bapat%
\ \BBA{} Beg%
}{%
{\protect\APACyear{1989}}%
}]{%
bapat1989order}%
\APACinsertmetastar{%
bapat1989order}%
Bapat, R\BPBI B.%
\BCBT{}\ \BBA{} Beg, M\BPBI I.%
%
\unskip\
\newblock
\APACrefYearMonthDay{1989}{}{}.
\newblock
\BBOQ{}\APACrefatitle{Order statistics for nonidentically distributed variables
  and permanents}{Order statistics for nonidentically distributed variables and
  permanents}.\BBCQ{}
\newblock
\APACjournalVolNumPages{Sankhy{\=a} Ser. A}{51}{}{79--93}.
\PrintBackRefs{\CurrentBib}

\bibitem[\protect\citeauthoryear{%
Barvinok%
}{%
Barvinok%
}{%
{\protect\APACyear{2010}}%
{\protect\APACexlab{{\protect\BCnt{1}}}}}]{%
barvinok2010matrices}%
\APACinsertmetastar{%
barvinok2010matrices}%
Barvinok, A.%
%
\unskip\
\newblock
\APACrefYearMonthDay{2010{\protect\BCnt{1}}}{}{}.
\newblock
\BBOQ{}\APACrefatitle{Matrices with prescribed row and column sums}{Matrices
  with prescribed row and column sums}.\BBCQ{}
\newblock
\APACjournalVolNumPages{Linear Algebra Appl.}{436}{}{820--834}.
\PrintBackRefs{\CurrentBib}

\bibitem[\protect\citeauthoryear{%
Barvinok%
}{%
Barvinok%
}{%
{\protect\APACyear{2010}}%
{\protect\APACexlab{{\protect\BCnt{2}}}}}]{%
barvinok2010number}%
\APACinsertmetastar{%
barvinok2010number}%
Barvinok, A.%
%
\unskip\
\newblock
\APACrefYearMonthDay{2010{\protect\BCnt{2}}}{}{}.
\newblock
\BBOQ{}\APACrefatitle{On the number of matrices and a random matrix with
  prescribed row and column sums and 0-1 entries}{On the number of matrices and
  a random matrix with prescribed row and column sums and 0-1 entries}.\BBCQ{}
\newblock
\APACjournalVolNumPages{Adv. Math.}{224}{1}{316--339}.
\PrintBackRefs{\CurrentBib}

\bibitem[\protect\citeauthoryear{%
Bayati%
, Kim%
\BCBL{}\ \BBA{} Saberi%
}{%
Bayati%
\ \protect\BOthers{.}}{%
{\protect\APACyear{2010}}%
}]{%
Bayati:Sequential:2009}%
\APACinsertmetastar{%
Bayati:Sequential:2009}%
Bayati, M.%
, Kim, J\BPBI H.%
\BCBL{}\ \BBA{} Saberi, A.%
%
\unskip\
\newblock
\APACrefYearMonthDay{2010}{}{}.
\newblock
\BBOQ{}\APACrefatitle{A sequential algorithm for generating random graphs}{A
  sequential algorithm for generating random graphs}.\BBCQ{}
\newblock
\APACjournalVolNumPages{Algorithmica}{58}{4}{860--910}.
\PrintBackRefs{\CurrentBib}

\bibitem[\protect\citeauthoryear{%
Beichl%
\ \BBA{} Sullivan%
}{%
Beichl%
\ \BBA{} Sullivan%
}{%
{\protect\APACyear{1999}}%
}]{%
beichl1999approximating}%
\APACinsertmetastar{%
beichl1999approximating}%
Beichl, I.%
\BCBT{}\ \BBA{} Sullivan, F.%
%
\unskip\
\newblock
\APACrefYearMonthDay{1999}{}{}.
\newblock
\BBOQ{}\APACrefatitle{Approximating the permanent via importance sampling with
  application to the dimer covering problem}{Approximating the permanent via
  importance sampling with application to the dimer covering problem}.\BBCQ{}
\newblock
\APACjournalVolNumPages{J. Comput. Phys.}{149}{1}{128--147}.
\PrintBackRefs{\CurrentBib}

\bibitem[\protect\citeauthoryear{%
B{\'e}k{\'e}ssy%
, Bekessy%
\BCBL{}\ \BBA{} Koml{\'o}s%
}{%
B{\'e}k{\'e}ssy%
\ \protect\BOthers{.}}{%
{\protect\APACyear{1972}}%
}]{%
bekessy1972asymptotic}%
\APACinsertmetastar{%
bekessy1972asymptotic}%
B{\'e}k{\'e}ssy, A.%
, Bekessy, P.%
\BCBL{}\ \BBA{} Koml{\'o}s, J.%
%
\unskip\
\newblock
\APACrefYearMonthDay{1972}{}{}.
\newblock
\BBOQ{}\APACrefatitle{Asymptotic enumeration of regular matrices}{Asymptotic
  enumeration of regular matrices}.\BBCQ{}
\newblock
\APACjournalVolNumPages{Stud. Sci. Math. Hungar.}{7}{}{343--353}.
\PrintBackRefs{\CurrentBib}

\bibitem[\protect\citeauthoryear{%
Besag%
\ \BBA{} Clifford%
}{%
Besag%
\ \BBA{} Clifford%
}{%
{\protect\APACyear{1989}}%
}]{%
besag1989gmc}%
\APACinsertmetastar{%
besag1989gmc}%
Besag, J.%
\BCBT{}\ \BBA{} Clifford, P.%
%
\unskip\
\newblock
\APACrefYearMonthDay{1989}{}{}.
\newblock
\BBOQ{}\APACrefatitle{{Generalized {M}onte {C}arlo significance
  tests}}{{Generalized {M}onte {C}arlo significance tests}}.\BBCQ{}
\newblock
\APACjournalVolNumPages{Biometrika}{76}{4}{633--642}.
\PrintBackRefs{\CurrentBib}

\bibitem[\protect\citeauthoryear{%
Bez\'akov\'a%
, Bhatnagar%
\BCBL{}\ \BBA{} Vigoda%
}{%
Bez\'akov\'a%
\ \protect\BOthers{.}}{%
{\protect\APACyear{2007}}%
}]{%
bezakova2007sbc}%
\APACinsertmetastar{%
bezakova2007sbc}%
Bez\'akov\'a, I.%
, Bhatnagar, N.%
\BCBL{}\ \BBA{} Vigoda, E.%
%
\unskip\
\newblock
\APACrefYearMonthDay{2007}{}{}.
\newblock
\BBOQ{}\APACrefatitle{{Sampling binary contingency tables with a greedy
  start}}{{Sampling binary contingency tables with a greedy start}}.\BBCQ{}
\newblock
\APACjournalVolNumPages{Random Struct. Algor.}{30}{}{168--205}.
\PrintBackRefs{\CurrentBib}

\bibitem[\protect\citeauthoryear{%
Bez\'akov\'a%
, Sinclair%
, \v{S}tefankovi\v{c}%
\BCBL{}\ \BBA{} Vigoda%
}{%
Bez\'akov\'a%
\ \protect\BOthers{.}}{%
{\protect\APACyear{2006}}%
}]{%
bezakova2006nes}%
\APACinsertmetastar{%
bezakova2006nes}%
Bez\'akov\'a, I.%
, Sinclair, A.%
, \v{S}tefankovi\v{c}, D.%
\BCBL{}\ \BBA{} Vigoda, E.%
%
\unskip\
\newblock
\APACrefYearMonthDay{2006}{}{}.
\newblock
\BBOQ{}\APACrefatitle{Negative Examples for Sequential Importance Sampling of
  Binary Contingency Tables}{Negative examples for sequential importance
  sampling of binary contingency tables}.\BBCQ{}
\newblock
\BIn{} Y.~Azar\ \BBA{} T.~Erlebach\ (\BEDS), \APACrefbtitle{Algorithms -- {ESA}
  2006}{Algorithms -- {ESA} 2006}\ (\BVOL\ 4168, \BPG~136-147).
\newblock
\APACaddressPublisher{Berlin/Heidelberg}{Springer}.
\PrintBackRefs{\CurrentBib}

\bibitem[\protect\citeauthoryear{%
Blanchet%
}{%
Blanchet%
}{%
{\protect\APACyear{2009}}%
}]{%
blanchet2006isa}%
\APACinsertmetastar{%
blanchet2006isa}%
Blanchet, J\BPBI H.%
%
\unskip\
\newblock
\APACrefYearMonthDay{2009}{}{}.
\newblock
\BBOQ{}\APACrefatitle{Efficient importance sampling for binary contingency
  tables}{Efficient importance sampling for binary contingency tables}.\BBCQ{}
\newblock
\APACjournalVolNumPages{Ann. Appl. Probab.}{19}{3}{949--982}.
\PrintBackRefs{\CurrentBib}

\bibitem[\protect\citeauthoryear{%
Brazzale%
}{%
Brazzale%
}{%
{\protect\APACyear{2005}}%
}]{%
cond}%
\APACinsertmetastar{%
cond}%
Brazzale, A\BPBI R.%
%
\unskip\
\newblock
\APACrefYearMonthDay{2005}{}{}.
\newblock
\BBOQ{}\APACrefatitle{{hoa}: {An} {R} package bundle for higher order
  likelihood inference}{{hoa}: {An} {R} package bundle for higher order
  likelihood inference}.\BBCQ{}
\newblock
\APACjournalVolNumPages{Rnews}{5}{}{20--27}.
\newblock
\APACrefnote{{ISSN} 609-3631}
\PrintBackRefs{\CurrentBib}

\bibitem[\protect\citeauthoryear{%
Brazzale%
\ \BBA{} Davison%
}{%
Brazzale%
\ \BBA{} Davison%
}{%
{\protect\APACyear{2008}}%
}]{%
brazzale2008accurate}%
\APACinsertmetastar{%
brazzale2008accurate}%
Brazzale, A\BPBI R.%
\BCBT{}\ \BBA{} Davison, A\BPBI C.%
%
\unskip\
\newblock
\APACrefYearMonthDay{2008}{}{}.
\newblock
\BBOQ{}\APACrefatitle{Accurate parametric inference for small samples}{Accurate
  parametric inference for small samples}.\BBCQ{}
\newblock
\APACjournalVolNumPages{Statist.~Sci.}{23}{4}{465--484}.
\PrintBackRefs{\CurrentBib}

\bibitem[\protect\citeauthoryear{%
Canfield%
, Greenhill%
\BCBL{}\ \BBA{} McKay%
}{%
Canfield%
\ \protect\BOthers{.}}{%
{\protect\APACyear{2008}}%
}]{%
canfield2008aed}%
\APACinsertmetastar{%
canfield2008aed}%
Canfield, E\BPBI R.%
, Greenhill, C.%
\BCBL{}\ \BBA{} McKay, B\BPBI D.%
%
\unskip\
\newblock
\APACrefYearMonthDay{2008}{}{}.
\newblock
\BBOQ{}\APACrefatitle{{Asymptotic enumeration of dense 0--1 matrices with
  specified line sums}}{{Asymptotic enumeration of dense 0--1 matrices with
  specified line sums}}.\BBCQ{}
\newblock
\APACjournalVolNumPages{J. Comb. Theory A}{115}{1}{32--66}.
\PrintBackRefs{\CurrentBib}

\bibitem[\protect\citeauthoryear{%
Chen%
}{%
Chen%
}{%
{\protect\APACyear{2006}}%
}]{%
chen2006sec}%
\APACinsertmetastar{%
chen2006sec}%
Chen, Y.%
%
\unskip\
\newblock
\APACrefYearMonthDay{2006}{}{}.
\newblock
\BBOQ{}\APACrefatitle{{Simple existence conditions for zero-one matrices with
  at most one structural zero in each row and column}}{{Simple existence
  conditions for zero-one matrices with at most one structural zero in each row
  and column}}.\BBCQ{}
\newblock
\APACjournalVolNumPages{Discrete Math.}{306}{22}{2870--2877}.
\PrintBackRefs{\CurrentBib}

\bibitem[\protect\citeauthoryear{%
Chen%
}{%
Chen%
}{%
{\protect\APACyear{2007}}%
}]{%
chen2007cit}%
\APACinsertmetastar{%
chen2007cit}%
Chen, Y.%
%
\unskip\
\newblock
\APACrefYearMonthDay{2007}{}{}.
\newblock
\BBOQ{}\APACrefatitle{{Conditional inference on tables with structural
  zeros}}{{Conditional inference on tables with structural zeros}}.\BBCQ{}
\newblock
\APACjournalVolNumPages{J. Comput. Graph. Stat.}{16}{2}{445--467}.
\PrintBackRefs{\CurrentBib}

\bibitem[\protect\citeauthoryear{%
Chen%
, Diaconis%
, Holmes%
\BCBL{}\ \BBA{} Liu%
}{%
Chen%
\ \protect\BOthers{.}}{%
{\protect\APACyear{2005}}%
}]{%
Chen:Sequential:2005}%
\APACinsertmetastar{%
Chen:Sequential:2005}%
Chen, Y.%
, Diaconis, P.%
, Holmes, S\BPBI P.%
\BCBL{}\ \BBA{} Liu, J\BPBI S.%
%
\unskip\
\newblock
\APACrefYearMonthDay{2005}{}{}.
\newblock
\BBOQ{}\APACrefatitle{Sequential {M}onte {C}arlo Methods for Statistical
  Analysis of Tables}{Sequential {M}onte {C}arlo methods for statistical
  analysis of tables}.\BBCQ{}
\newblock
\APACjournalVolNumPages{J. Am. Statist. Assoc.}{100}{469}{109--120}.
\PrintBackRefs{\CurrentBib}

\bibitem[\protect\citeauthoryear{%
Chen%
\ \BBA{} Small%
}{%
Chen%
\ \BBA{} Small%
}{%
{\protect\APACyear{2005}}%
}]{%
chen2005exact}%
\APACinsertmetastar{%
chen2005exact}%
Chen, Y.%
\BCBT{}\ \BBA{} Small, D.%
%
\unskip\
\newblock
\APACrefYearMonthDay{2005}{}{}.
\newblock
\BBOQ{}\APACrefatitle{Exact tests for the {R}asch model via sequential
  importance sampling}{Exact tests for the {R}asch model via sequential
  importance sampling}.\BBCQ{}
\newblock
\APACjournalVolNumPages{Psychometrika}{70}{1}{11--30}.
\PrintBackRefs{\CurrentBib}

\bibitem[\protect\citeauthoryear{%
Connor%
\ \BBA{} Simberloff%
}{%
Connor%
\ \BBA{} Simberloff%
}{%
{\protect\APACyear{1979}}%
}]{%
connor1979assembly}%
\APACinsertmetastar{%
connor1979assembly}%
Connor, E\BPBI F.%
\BCBT{}\ \BBA{} Simberloff, D.%
%
\unskip\
\newblock
\APACrefYearMonthDay{1979}{}{}.
\newblock
\BBOQ{}\APACrefatitle{The assembly of species communities: chance or
  competition?}{The assembly of species communities: chance or
  competition?}\BBCQ{}
\newblock
\APACjournalVolNumPages{Ecology}{60}{}{1132--1140}.
\PrintBackRefs{\CurrentBib}

\bibitem[\protect\citeauthoryear{%
Cox%
}{%
Cox%
}{%
{\protect\APACyear{1958}}%
}]{%
cox1958regression}%
\APACinsertmetastar{%
cox1958regression}%
Cox, D\BPBI R.%
%
\unskip\
\newblock
\APACrefYearMonthDay{1958}{}{}.
\newblock
\BBOQ{}\APACrefatitle{The regression analysis of binary sequences}{The
  regression analysis of binary sequences}.\BBCQ{}
\newblock
\APACjournalVolNumPages{J. R. Statist. Soc. B}{20}{}{215--242}.
\PrintBackRefs{\CurrentBib}

\bibitem[\protect\citeauthoryear{%
Cytel%
}{%
Cytel%
}{%
{\protect\APACyear{2010}}%
}]{%
logxact}%
\APACinsertmetastar{%
logxact}%
Cytel.%
%
\unskip\
\newblock
\APACrefYear{2010}.
\newblock
\APACrefbtitle{Log{X}act 9}{Log{X}act 9}.
\newblock
\APACaddressPublisher{Cambridge, MA}{Cytel Inc}.
\PrintBackRefs{\CurrentBib}

\bibitem[\protect\citeauthoryear{%
Diaconis%
\ \BBA{} Evans%
}{%
Diaconis%
\ \BBA{} Evans%
}{%
{\protect\APACyear{2000}}%
}]{%
diaconis2000immanants}%
\APACinsertmetastar{%
diaconis2000immanants}%
Diaconis, P.%
\BCBT{}\ \BBA{} Evans, S\BPBI N.%
%
\unskip\
\newblock
\APACrefYearMonthDay{2000}{}{}.
\newblock
\BBOQ{}\APACrefatitle{Immanants and finite point processes}{Immanants and
  finite point processes}.\BBCQ{}
\newblock
\APACjournalVolNumPages{J. Comb. Theory A}{91}{1-2}{305--321}.
\PrintBackRefs{\CurrentBib}

\bibitem[\protect\citeauthoryear{%
Durrett%
}{%
Durrett%
}{%
{\protect\APACyear{2010}}%
}]{%
durrett2010probability}%
\APACinsertmetastar{%
durrett2010probability}%
Durrett, R.%
%
\unskip\
\newblock
\APACrefYear{2010}.
\newblock
\APACrefbtitle{Probability: theory and examples}{Probability: theory and
  examples}\ (\PrintOrdinal{4th}\ \BEd).
\newblock
\APACaddressPublisher{New York}{Cambridge Univ. Pr.}
\PrintBackRefs{\CurrentBib}

\bibitem[\protect\citeauthoryear{%
Fienberg%
, Meyer%
\BCBL{}\ \BBA{} Wasserman%
}{%
Fienberg%
\ \protect\BOthers{.}}{%
{\protect\APACyear{1985}}%
}]{%
fienberg1985statistical}%
\APACinsertmetastar{%
fienberg1985statistical}%
Fienberg, S.%
, Meyer, M.%
\BCBL{}\ \BBA{} Wasserman, S.%
%
\unskip\
\newblock
\APACrefYearMonthDay{1985}{}{}.
\newblock
\BBOQ{}\APACrefatitle{Statistical analysis of multiple sociometric
  relations}{Statistical analysis of multiple sociometric relations}.\BBCQ{}
\newblock
\APACjournalVolNumPages{Journal of the American Statistical
  Association}{80}{389}{51--67}.
\PrintBackRefs{\CurrentBib}

\bibitem[\protect\citeauthoryear{%
Frey%
}{%
Frey%
}{%
{\protect\APACyear{1998}}%
}]{%
Frey:Graphical:1998}%
\APACinsertmetastar{%
Frey:Graphical:1998}%
Frey, B\BPBI J.%
%
\unskip\
\newblock
\APACrefYear{1998}.
\newblock
\APACrefbtitle{Graphical Models for Machine Learning and Digital
  Communication}{Graphical models for machine learning and digital
  communication}.
\newblock
\APACaddressPublisher{Cambridge, MA}{MIT Press}.
\PrintBackRefs{\CurrentBib}

\bibitem[\protect\citeauthoryear{%
Gale%
}{%
Gale%
}{%
{\protect\APACyear{1957}}%
}]{%
gale1957tfn}%
\APACinsertmetastar{%
gale1957tfn}%
Gale, D.%
%
\unskip\
\newblock
\APACrefYearMonthDay{1957}{}{}.
\newblock
\BBOQ{}\APACrefatitle{{A theorem on flows in networks}}{{A theorem on flows in
  networks}}.\BBCQ{}
\newblock
\APACjournalVolNumPages{Pac. J. Math.}{7}{}{1073--1082}.
\PrintBackRefs{\CurrentBib}

\bibitem[\protect\citeauthoryear{%
Goldenberg%
, Zheng%
, Fienberg%
\BCBL{}\ \BBA{} Airoldi%
}{%
Goldenberg%
\ \protect\BOthers{.}}{%
{\protect\APACyear{2010}}%
}]{%
goldenberg2010survey}%
\APACinsertmetastar{%
goldenberg2010survey}%
Goldenberg, A.%
, Zheng, A.%
, Fienberg, S.%
\BCBL{}\ \BBA{} Airoldi, E.%
%
\unskip\
\newblock
\APACrefYearMonthDay{2010}{}{}.
\newblock
\BBOQ{}\APACrefatitle{A survey of statistical network models}{A survey of
  statistical network models}.\BBCQ{}
\newblock
\APACjournalVolNumPages{Foundations and Trends in Machine
  Learning}{2}{2}{129--233}.
\PrintBackRefs{\CurrentBib}

\bibitem[\protect\citeauthoryear{%
Gotelli%
}{%
Gotelli%
}{%
{\protect\APACyear{2000}}%
}]{%
gotelli2000null}%
\APACinsertmetastar{%
gotelli2000null}%
Gotelli, N\BPBI J.%
%
\unskip\
\newblock
\APACrefYearMonthDay{2000}{}{}.
\newblock
\BBOQ{}\APACrefatitle{Null model analysis of species co-occurrence
  patterns}{Null model analysis of species co-occurrence patterns}.\BBCQ{}
\newblock
\APACjournalVolNumPages{Ecology}{81}{9}{2606--2621}.
\PrintBackRefs{\CurrentBib}

\bibitem[\protect\citeauthoryear{%
Greenhill%
\ \BBA{} McKay%
}{%
Greenhill%
\ \BBA{} McKay%
}{%
{\protect\APACyear{2009}}%
}]{%
greenhill2009random}%
\APACinsertmetastar{%
greenhill2009random}%
Greenhill, C.%
\BCBT{}\ \BBA{} McKay, B\BPBI D.%
%
\unskip\
\newblock
\APACrefYearMonthDay{2009}{}{}.
\newblock
\BBOQ{}\APACrefatitle{Random dense bipartite graphs and directed graphs with
  specified degrees}{Random dense bipartite graphs and directed graphs with
  specified degrees}.\BBCQ{}
\newblock
\APACjournalVolNumPages{Random Struct. Algor.}{35}{2}{222--249}.
\PrintBackRefs{\CurrentBib}

\bibitem[\protect\citeauthoryear{%
Greenhill%
, McKay%
\BCBL{}\ \BBA{} Wang%
}{%
Greenhill%
\ \protect\BOthers{.}}{%
{\protect\APACyear{2006}}%
}]{%
greenhill2006aes}%
\APACinsertmetastar{%
greenhill2006aes}%
Greenhill, C.%
, McKay, B\BPBI D.%
\BCBL{}\ \BBA{} Wang, X.%
%
\unskip\
\newblock
\APACrefYearMonthDay{2006}{}{}.
\newblock
\BBOQ{}\APACrefatitle{{Asymptotic enumeration of sparse 0--1 matrices with
  irregular row and column sums}}{{Asymptotic enumeration of sparse 0--1
  matrices with irregular row and column sums}}.\BBCQ{}
\newblock
\APACjournalVolNumPages{J. Comb. Theory A}{113}{2}{291--324}.
\PrintBackRefs{\CurrentBib}

\bibitem[\protect\citeauthoryear{%
Harrison%
}{%
Harrison%
}{%
{\protect\APACyear{2012}}%
}]{%
harrison2012conservative}%
\APACinsertmetastar{%
harrison2012conservative}%
Harrison, M\BPBI T.%
%
\unskip\
\newblock
\APACrefYearMonthDay{2012}{}{}.
\newblock
\BBOQ{}\APACrefatitle{Conservative hypothesis tests and confidence intervals
  using importance sampling}{Conservative hypothesis tests and confidence
  intervals using importance sampling}.\BBCQ{}
\newblock
\APACjournalVolNumPages{Biometrika}{99}{1}{57--69}.
\PrintBackRefs{\CurrentBib}

\bibitem[\protect\citeauthoryear{%
Harrison%
\ \BBA{} Geman%
}{%
Harrison%
\ \BBA{} Geman%
}{%
{\protect\APACyear{2009}}%
}]{%
harrison2009rate}%
\APACinsertmetastar{%
harrison2009rate}%
Harrison, M\BPBI T.%
\BCBT{}\ \BBA{} Geman, S.%
%
\unskip\
\newblock
\APACrefYearMonthDay{2009}{}{}.
\newblock
\BBOQ{}\APACrefatitle{A rate and history-preserving resampling algorithm for
  neural spike trains}{A rate and history-preserving resampling algorithm for
  neural spike trains}.\BBCQ{}
\newblock
\APACjournalVolNumPages{Neural comput.}{21}{5}{1244--1258}.
\PrintBackRefs{\CurrentBib}

\bibitem[\protect\citeauthoryear{%
Holland%
\ \BBA{} Leinhardt%
}{%
Holland%
\ \BBA{} Leinhardt%
}{%
{\protect\APACyear{1981}}%
}]{%
holland1981exponential}%
\APACinsertmetastar{%
holland1981exponential}%
Holland, P\BPBI W.%
\BCBT{}\ \BBA{} Leinhardt, S.%
%
\unskip\
\newblock
\APACrefYearMonthDay{1981}{}{}.
\newblock
\BBOQ{}\APACrefatitle{An exponential family of probability distributions for
  directed graphs}{An exponential family of probability distributions for
  directed graphs}.\BBCQ{}
\newblock
\APACjournalVolNumPages{J. Am. Statist. Assoc.}{76}{}{33--50}.
\PrintBackRefs{\CurrentBib}

\bibitem[\protect\citeauthoryear{%
Jerrum%
, Sinclair%
\BCBL{}\ \BBA{} Vigoda%
}{%
Jerrum%
\ \protect\BOthers{.}}{%
{\protect\APACyear{2004}}%
}]{%
jerrum2004polynomial}%
\APACinsertmetastar{%
jerrum2004polynomial}%
Jerrum, M.%
, Sinclair, A.%
\BCBL{}\ \BBA{} Vigoda, E.%
%
\unskip\
\newblock
\APACrefYearMonthDay{2004}{}{}.
\newblock
\BBOQ{}\APACrefatitle{A polynomial-time approximation algorithm for the
  permanent of a matrix with nonnegative entries}{A polynomial-time
  approximation algorithm for the permanent of a matrix with nonnegative
  entries}.\BBCQ{}
\newblock
\APACjournalVolNumPages{J. Assoc. Comp. Mach.}{51}{4}{671--697}.
\PrintBackRefs{\CurrentBib}

\bibitem[\protect\citeauthoryear{%
Kannan%
, Tetali%
\BCBL{}\ \BBA{} Vempala%
}{%
Kannan%
\ \protect\BOthers{.}}{%
{\protect\APACyear{1999}}%
}]{%
kannan1999simple}%
\APACinsertmetastar{%
kannan1999simple}%
Kannan, R.%
, Tetali, P.%
\BCBL{}\ \BBA{} Vempala, S.%
%
\unskip\
\newblock
\APACrefYearMonthDay{1999}{}{}.
\newblock
\BBOQ{}\APACrefatitle{Simple {M}arkov-chain algorithms for generating bipartite
  graphs and tournaments}{Simple {M}arkov-chain algorithms for generating
  bipartite graphs and tournaments}.\BBCQ{}
\newblock
\APACjournalVolNumPages{Random Struct. Algor.}{14}{4}{293--308}.
\PrintBackRefs{\CurrentBib}

\bibitem[\protect\citeauthoryear{%
Kong%
, Liu%
\BCBL{}\ \BBA{} Wong%
}{%
Kong%
\ \protect\BOthers{.}}{%
{\protect\APACyear{1994}}%
}]{%
Kong:Sequential:1994}%
\APACinsertmetastar{%
Kong:Sequential:1994}%
Kong, A.%
, Liu, J\BPBI S.%
\BCBL{}\ \BBA{} Wong, W\BPBI H.%
%
\unskip\
\newblock
\APACrefYearMonthDay{1994}{}{}.
\newblock
\BBOQ{}\APACrefatitle{Sequential imputations and {B}ayesian missing data
  problems}{Sequential imputations and {B}ayesian missing data
  problems}.\BBCQ{}
\newblock
\APACjournalVolNumPages{J. Am. Statist. Assoc.}{89}{}{278--288}.
\PrintBackRefs{\CurrentBib}

\bibitem[\protect\citeauthoryear{%
Kou%
\ \BBA{} McCullagh%
}{%
Kou%
\ \BBA{} McCullagh%
}{%
{\protect\APACyear{2009}}%
}]{%
kou2009approximating}%
\APACinsertmetastar{%
kou2009approximating}%
Kou, S\BPBI C.%
\BCBT{}\ \BBA{} McCullagh, P.%
%
\unskip\
\newblock
\APACrefYearMonthDay{2009}{}{}.
\newblock
\BBOQ{}\APACrefatitle{Approximating the $\alpha$-permanent}{Approximating the
  $\alpha$-permanent}.\BBCQ{}
\newblock
\APACjournalVolNumPages{Biometrika}{96}{3}{635--644}.
\PrintBackRefs{\CurrentBib}

\bibitem[\protect\citeauthoryear{%
Littlewood%
}{%
Littlewood%
}{%
{\protect\APACyear{1950}}%
}]{%
littlewood1950theory}%
\APACinsertmetastar{%
littlewood1950theory}%
Littlewood, D.%
%
\unskip\
\newblock
\APACrefYear{1950}.
\newblock
\APACrefbtitle{The theory of group characters and matrix representations of
  groups}{The theory of group characters and matrix representations of groups}.
\newblock
\APACaddressPublisher{}{Oxford Univ. Press}.
\PrintBackRefs{\CurrentBib}

\bibitem[\protect\citeauthoryear{%
Liu%
}{%
Liu%
}{%
{\protect\APACyear{2001}}%
}]{%
Liu:Monte:2001}%
\APACinsertmetastar{%
Liu:Monte:2001}%
Liu, J\BPBI S.%
%
\unskip\
\newblock
\APACrefYear{2001}.
\newblock
\APACrefbtitle{{M}onte {C}arlo Strategies in Scientific Computing}{{M}onte
  {C}arlo strategies in scientific computing}.
\newblock
\APACaddressPublisher{New York}{Springer}.
\PrintBackRefs{\CurrentBib}

\bibitem[\protect\citeauthoryear{%
Macchi%
}{%
Macchi%
}{%
{\protect\APACyear{1975}}%
}]{%
macchi1975coincidence}%
\APACinsertmetastar{%
macchi1975coincidence}%
Macchi, O.%
%
\unskip\
\newblock
\APACrefYearMonthDay{1975}{}{}.
\newblock
\BBOQ{}\APACrefatitle{The coincidence approach to stochastic point
  processes}{The coincidence approach to stochastic point processes}.\BBCQ{}
\newblock
\APACjournalVolNumPages{Adv. Appl. Probab.}{7}{}{83--122}.
\PrintBackRefs{\CurrentBib}

\bibitem[\protect\citeauthoryear{%
Matsumoto%
\ \BBA{} Nishimura%
}{%
Matsumoto%
\ \BBA{} Nishimura%
}{%
{\protect\APACyear{1998}}%
}]{%
matsumoto1998mersenne}%
\APACinsertmetastar{%
matsumoto1998mersenne}%
Matsumoto, M.%
\BCBT{}\ \BBA{} Nishimura, T.%
%
\unskip\
\newblock
\APACrefYearMonthDay{1998}{}{}.
\newblock
\BBOQ{}\APACrefatitle{Mersenne twister: a 623-dimensionally equidistributed
  uniform pseudo-random number generator}{Mersenne twister: a 623-dimensionally
  equidistributed uniform pseudo-random number generator}.\BBCQ{}
\newblock
\APACjournalVolNumPages{ACM Trans. Model. Comp. Simul.}{8}{1}{3--30}.
\PrintBackRefs{\CurrentBib}

\bibitem[\protect\citeauthoryear{%
McCullagh%
\ \BBA{} M{\o}ller%
}{%
McCullagh%
\ \BBA{} M{\o}ller%
}{%
{\protect\APACyear{2006}}%
}]{%
mccullagh2006permanental}%
\APACinsertmetastar{%
mccullagh2006permanental}%
McCullagh, P.%
\BCBT{}\ \BBA{} M{\o}ller, J.%
%
\unskip\
\newblock
\APACrefYearMonthDay{2006}{}{}.
\newblock
\BBOQ{}\APACrefatitle{The permanental process}{The permanental process}.\BBCQ{}
\newblock
\APACjournalVolNumPages{Adv. Appl. Probab.}{38}{}{873--888}.
\PrintBackRefs{\CurrentBib}

\bibitem[\protect\citeauthoryear{%
McKay%
}{%
McKay%
}{%
{\protect\APACyear{1984}}%
}]{%
mckay1984amp}%
\APACinsertmetastar{%
mckay1984amp}%
McKay, B\BPBI D.%
%
\unskip\
\newblock
\APACrefYearMonthDay{1984}{}{}.
\newblock
\BBOQ{}\APACrefatitle{Asymptotics for 0-1 matrices with prescribed line
  sums}{Asymptotics for 0-1 matrices with prescribed line sums}.\BBCQ{}
\newblock
\BIn{} D\BPBI M.~Jackson\ \BBA{} S\BPBI A.~Vanstone\ (\BEDS),
  \APACrefbtitle{Enumeration and Design}{Enumeration and design}\ (\BPGS\
  225--238).
\newblock
\APACaddressPublisher{}{Academic Press}.
\PrintBackRefs{\CurrentBib}

\bibitem[\protect\citeauthoryear{%
McKay%
\ \BBA{} Wormald%
}{%
McKay%
\ \BBA{} Wormald%
}{%
{\protect\APACyear{1990}}%
}]{%
mckay1990uniform}%
\APACinsertmetastar{%
mckay1990uniform}%
McKay, B\BPBI D.%
\BCBT{}\ \BBA{} Wormald, N\BPBI C.%
%
\unskip\
\newblock
\APACrefYearMonthDay{1990}{}{}.
\newblock
\BBOQ{}\APACrefatitle{Uniform generation of random regular graphs of moderate
  degree}{Uniform generation of random regular graphs of moderate
  degree}.\BBCQ{}
\newblock
\APACjournalVolNumPages{Journal of Algorithms}{11}{1}{52--67}.
\PrintBackRefs{\CurrentBib}

\bibitem[\protect\citeauthoryear{%
Mehta%
\ \BBA{} Patel%
}{%
Mehta%
\ \BBA{} Patel%
}{%
{\protect\APACyear{1995}}%
}]{%
mehta1995exact}%
\APACinsertmetastar{%
mehta1995exact}%
Mehta, C\BPBI R.%
\BCBT{}\ \BBA{} Patel, N\BPBI R.%
%
\unskip\
\newblock
\APACrefYearMonthDay{1995}{}{}.
\newblock
\BBOQ{}\APACrefatitle{Exact logistic regression: theory and examples}{Exact
  logistic regression: theory and examples}.\BBCQ{}
\newblock
\APACjournalVolNumPages{Statist. Med.}{14}{19}{2143--2160}.
\PrintBackRefs{\CurrentBib}

\bibitem[\protect\citeauthoryear{%
O'Neil%
}{%
O'Neil%
}{%
{\protect\APACyear{1969}}%
}]{%
Oneil:Asymptotics:1969}%
\APACinsertmetastar{%
Oneil:Asymptotics:1969}%
O'Neil, P\BPBI E.%
%
\unskip\
\newblock
\APACrefYearMonthDay{1969}{}{}.
\newblock
\BBOQ{}\APACrefatitle{Asymptotics and random matrices with row-sum and
  column-sum restrictions}{Asymptotics and random matrices with row-sum and
  column-sum restrictions}.\BBCQ{}
\newblock
\APACjournalVolNumPages{B. Am. Math. Soc.}{75}{}{1276--1282}.
\PrintBackRefs{\CurrentBib}

\bibitem[\protect\citeauthoryear{%
Park%
\ \BBA{} Miller%
}{%
Park%
\ \BBA{} Miller%
}{%
{\protect\APACyear{1988}}%
}]{%
park1988random}%
\APACinsertmetastar{%
park1988random}%
Park, S\BPBI K.%
\BCBT{}\ \BBA{} Miller, K\BPBI W.%
%
\unskip\
\newblock
\APACrefYearMonthDay{1988}{}{}.
\newblock
\BBOQ{}\APACrefatitle{Random number generators: good ones are hard to
  find}{Random number generators: good ones are hard to find}.\BBCQ{}
\newblock
\APACjournalVolNumPages{Commun. Assoc. Comp. Mach.}{31}{10}{1192--1201}.
\PrintBackRefs{\CurrentBib}

\bibitem[\protect\citeauthoryear{%
Ponocny%
}{%
Ponocny%
}{%
{\protect\APACyear{2001}}%
}]{%
ponocny2001nonparametric}%
\APACinsertmetastar{%
ponocny2001nonparametric}%
Ponocny, I.%
%
\unskip\
\newblock
\APACrefYearMonthDay{2001}{}{}.
\newblock
\BBOQ{}\APACrefatitle{Nonparametric goodness-of-fit tests for the {R}asch
  model}{Nonparametric goodness-of-fit tests for the {R}asch model}.\BBCQ{}
\newblock
\APACjournalVolNumPages{Psychometrika}{66}{3}{437--459}.
\PrintBackRefs{\CurrentBib}

\bibitem[\protect\citeauthoryear{%
Rao%
, Jana%
\BCBL{}\ \BBA{} Bandyopadhyay%
}{%
Rao%
\ \protect\BOthers{.}}{%
{\protect\APACyear{1996}}%
}]{%
rao1996markov}%
\APACinsertmetastar{%
rao1996markov}%
Rao, A.%
, Jana, R.%
\BCBL{}\ \BBA{} Bandyopadhyay, S.%
%
\unskip\
\newblock
\APACrefYearMonthDay{1996}{}{}.
\newblock
\BBOQ{}\APACrefatitle{A {M}arkov chain {M}onte {C}arlo method for generating
  random (0, 1)-matrices with given marginals}{A {M}arkov chain {M}onte {C}arlo
  method for generating random (0, 1)-matrices with given marginals}.\BBCQ{}
\newblock
\APACjournalVolNumPages{Sankhy{\=a}: The Indian Journal of Statistics, Series
  A}{}{}{225--242}.
\PrintBackRefs{\CurrentBib}

\bibitem[\protect\citeauthoryear{%
Rasch%
}{%
Rasch%
}{%
{\protect\APACyear{1960}}%
}]{%
rasch1960probabilistic}%
\APACinsertmetastar{%
rasch1960probabilistic}%
Rasch, G.%
%
\unskip\
\newblock
\APACrefYear{1960}.
\newblock
\APACrefbtitle{Probabilistic models for some intelligence and attainment
  tests}{Probabilistic models for some intelligence and attainment tests}.
\newblock
\APACaddressPublisher{Copenhagen}{Danmarks Paedagogiske Institut.}
\PrintBackRefs{\CurrentBib}

\bibitem[\protect\citeauthoryear{%
Rasch%
}{%
Rasch%
}{%
{\protect\APACyear{1961}}%
}]{%
rasch1961general}%
\APACinsertmetastar{%
rasch1961general}%
Rasch, G.%
%
\unskip\
\newblock
\APACrefYearMonthDay{1961}{}{}.
\newblock
\BBOQ{}\APACrefatitle{{On general laws and the meaning of measurement in
  psychology}}{{On general laws and the meaning of measurement in
  psychology}}.\BBCQ{}
\newblock
\BIn{} J.~Neyman\ (\BED), \APACrefbtitle{Proceedings of the Fourth Berkeley
  Symposium on Mathematical Statistics and Probability: Probability
  theory}{Proceedings of the fourth berkeley symposium on mathematical
  statistics and probability: Probability theory}\ (\BVOL~4, \BPGS\ 321--334).
\newblock
\APACaddressPublisher{Berkeley, CA}{}.
\PrintBackRefs{\CurrentBib}

\bibitem[\protect\citeauthoryear{%
Rothblum%
\ \BBA{} Schneider%
}{%
Rothblum%
\ \BBA{} Schneider%
}{%
{\protect\APACyear{1989}}%
}]{%
rothblum1989scalings}%
\APACinsertmetastar{%
rothblum1989scalings}%
Rothblum, U\BPBI G.%
\BCBT{}\ \BBA{} Schneider, H.%
%
\unskip\
\newblock
\APACrefYearMonthDay{1989}{}{}.
\newblock
\BBOQ{}\APACrefatitle{Scalings of matrices which have prespecified row sums and
  column sums via optimization}{Scalings of matrices which have prespecified
  row sums and column sums via optimization}.\BBCQ{}
\newblock
\APACjournalVolNumPages{Linear Algebra Appl.}{114}{}{737--764}.
\PrintBackRefs{\CurrentBib}

\bibitem[\protect\citeauthoryear{%
Ryser%
}{%
Ryser%
}{%
{\protect\APACyear{1957}}%
}]{%
ryser1957cpm}%
\APACinsertmetastar{%
ryser1957cpm}%
Ryser, H\BPBI J.%
%
\unskip\
\newblock
\APACrefYearMonthDay{1957}{}{}.
\newblock
\BBOQ{}\APACrefatitle{Combinatorial properties of matrices of zeros and
  ones}{Combinatorial properties of matrices of zeros and ones}.\BBCQ{}
\newblock
\APACjournalVolNumPages{Can. J. Math.}{9}{}{371--377}.
\PrintBackRefs{\CurrentBib}

\bibitem[\protect\citeauthoryear{%
Shirai%
\ \BBA{} Takahashi%
}{%
Shirai%
\ \BBA{} Takahashi%
}{%
{\protect\APACyear{2003}}%
}]{%
shirai2003random}%
\APACinsertmetastar{%
shirai2003random}%
Shirai, T.%
\BCBT{}\ \BBA{} Takahashi, Y.%
%
\unskip\
\newblock
\APACrefYearMonthDay{2003}{}{}.
\newblock
\BBOQ{}\APACrefatitle{Random point fields associated with certain {F}redholm
  determinants {I}: fermion, {P}oisson and boson point processes}{Random point
  fields associated with certain {F}redholm determinants {I}: fermion,
  {P}oisson and boson point processes}.\BBCQ{}
\newblock
\APACjournalVolNumPages{J. Funct. Anal.}{205}{2}{414--463}.
\PrintBackRefs{\CurrentBib}

\bibitem[\protect\citeauthoryear{%
Sinkhorn%
}{%
Sinkhorn%
}{%
{\protect\APACyear{1964}}%
}]{%
sinkhorn1964relationship}%
\APACinsertmetastar{%
sinkhorn1964relationship}%
Sinkhorn, R.%
%
\unskip\
\newblock
\APACrefYearMonthDay{1964}{}{}.
\newblock
\BBOQ{}\APACrefatitle{A relationship between arbitrary positive matrices and
  doubly stochastic matrices}{A relationship between arbitrary positive
  matrices and doubly stochastic matrices}.\BBCQ{}
\newblock
\APACjournalVolNumPages{Ann. Math. Statist.}{35}{2}{876--879}.
\PrintBackRefs{\CurrentBib}

\bibitem[\protect\citeauthoryear{%
Sinkhorn%
}{%
Sinkhorn%
}{%
{\protect\APACyear{1967}}%
}]{%
sinkhorn1967diagonal}%
\APACinsertmetastar{%
sinkhorn1967diagonal}%
Sinkhorn, R.%
%
\unskip\
\newblock
\APACrefYearMonthDay{1967}{}{}.
\newblock
\BBOQ{}\APACrefatitle{Diagonal equivalence to matrices with prescribed row and
  column sums}{Diagonal equivalence to matrices with prescribed row and column
  sums}.\BBCQ{}
\newblock
\APACjournalVolNumPages{Am. Math. Mon.}{74}{4}{402--405}.
\PrintBackRefs{\CurrentBib}

\bibitem[\protect\citeauthoryear{%
Snijders%
}{%
Snijders%
}{%
{\protect\APACyear{1991}}%
}]{%
snijders1991enumeration}%
\APACinsertmetastar{%
snijders1991enumeration}%
Snijders, T\BPBI A\BPBI B.%
%
\unskip\
\newblock
\APACrefYearMonthDay{1991}{}{}.
\newblock
\BBOQ{}\APACrefatitle{Enumeration and simulation methods for 0--1 matrices with
  given marginals}{Enumeration and simulation methods for 0--1 matrices with
  given marginals}.\BBCQ{}
\newblock
\APACjournalVolNumPages{Psychometrika}{56}{3}{397--417}.
\PrintBackRefs{\CurrentBib}

\bibitem[\protect\citeauthoryear{%
StataCorp%
}{%
StataCorp%
}{%
{\protect\APACyear{2009}}%
}]{%
stata}%
\APACinsertmetastar{%
stata}%
StataCorp.%
%
\unskip\
\newblock
\APACrefYear{2009}.
\newblock
\APACrefbtitle{Stata Statistical Software: Release 11}{Stata statistical
  software: Release 11}.
\newblock
\APACaddressPublisher{College Station, TX}{StataCorp LP}.
\PrintBackRefs{\CurrentBib}

\bibitem[\protect\citeauthoryear{%
Valiant%
}{%
Valiant%
}{%
{\protect\APACyear{1979}}%
}]{%
valiant1979complexity}%
\APACinsertmetastar{%
valiant1979complexity}%
Valiant, L\BPBI G.%
%
\unskip\
\newblock
\APACrefYearMonthDay{1979}{}{}.
\newblock
\BBOQ{}\APACrefatitle{The complexity of computing the permanent}{The complexity
  of computing the permanent}.\BBCQ{}
\newblock
\APACjournalVolNumPages{Theor. Comput. Sci.}{8}{2}{189--201}.
\PrintBackRefs{\CurrentBib}

\bibitem[\protect\citeauthoryear{%
Vaughan%
\ \BBA{} Venables%
}{%
Vaughan%
\ \BBA{} Venables%
}{%
{\protect\APACyear{1972}}%
}]{%
vaughan1972permanent}%
\APACinsertmetastar{%
vaughan1972permanent}%
Vaughan, R\BPBI J.%
\BCBT{}\ \BBA{} Venables, W\BPBI N.%
%
\unskip\
\newblock
\APACrefYearMonthDay{1972}{}{}.
\newblock
\BBOQ{}\APACrefatitle{Permanent expressions for order statistic
  densities}{Permanent expressions for order statistic densities}.\BBCQ{}
\newblock
\APACjournalVolNumPages{J. R. Statist. Soc. B}{34}{}{308--310}.
\PrintBackRefs{\CurrentBib}

\bibitem[\protect\citeauthoryear{%
Vere-Jones%
}{%
Vere-Jones%
}{%
{\protect\APACyear{1988}}%
}]{%
vere1988generalization}%
\APACinsertmetastar{%
vere1988generalization}%
Vere-Jones, D.%
%
\unskip\
\newblock
\APACrefYearMonthDay{1988}{}{}.
\newblock
\BBOQ{}\APACrefatitle{A generalization of permanents and determinants}{A
  generalization of permanents and determinants}.\BBCQ{}
\newblock
\APACjournalVolNumPages{Linear Algebra Appl.}{111}{}{119--124}.
\PrintBackRefs{\CurrentBib}

\bibitem[\protect\citeauthoryear{%
Vere-Jones%
}{%
Vere-Jones%
}{%
{\protect\APACyear{1997}}%
}]{%
vere1997alpha}%
\APACinsertmetastar{%
vere1997alpha}%
Vere-Jones, D.%
%
\unskip\
\newblock
\APACrefYearMonthDay{1997}{}{}.
\newblock
\BBOQ{}\APACrefatitle{Alpha-permanents and their applications to multivariate
  gamma, negative binomial and ordinary binomial
  distributions}{Alpha-permanents and their applications to multivariate gamma,
  negative binomial and ordinary binomial distributions}.\BBCQ{}
\newblock
\APACjournalVolNumPages{New Zeal. J. Math.}{26}{}{125--149}.
\PrintBackRefs{\CurrentBib}

\bibitem[\protect\citeauthoryear{%
Verhelst%
}{%
Verhelst%
}{%
{\protect\APACyear{2008}}%
}]{%
verhelst2008ema}%
\APACinsertmetastar{%
verhelst2008ema}%
Verhelst, N\BPBI D.%
%
\unskip\
\newblock
\APACrefYearMonthDay{2008}{}{}.
\newblock
\BBOQ{}\APACrefatitle{An Efficient {MCMC} Algorithm to Sample Binary Matrices
  with Fixed Marginals}{An efficient {MCMC} algorithm to sample binary matrices
  with fixed marginals}.\BBCQ{}
\newblock
\APACjournalVolNumPages{Psychometrika}{73}{4}{705--728}.
\PrintBackRefs{\CurrentBib}

\bibitem[\protect\citeauthoryear{%
Wasserman%
}{%
Wasserman%
}{%
{\protect\APACyear{1977}}%
}]{%
wasserman1977random}%
\APACinsertmetastar{%
wasserman1977random}%
Wasserman, S\BPBI S.%
%
\unskip\
\newblock
\APACrefYearMonthDay{1977}{}{}.
\newblock
\BBOQ{}\APACrefatitle{Random directed graph distributions and the triad census
  in social networks }{Random directed graph distributions and the triad census
  in social networks }.\BBCQ{}
\newblock
\APACjournalVolNumPages{J. Math. Sociol.}{5}{1}{61--86}.
\PrintBackRefs{\CurrentBib}

\bibitem[\protect\citeauthoryear{%
Zamar%
, McNeney%
\BCBL{}\ \BBA{} Graham%
}{%
Zamar%
\ \protect\BOthers{.}}{%
{\protect\APACyear{2007}}%
}]{%
zamar2007elrm}%
\APACinsertmetastar{%
zamar2007elrm}%
Zamar, D.%
, McNeney, B.%
\BCBL{}\ \BBA{} Graham, J.%
%
\unskip\
\newblock
\APACrefYearMonthDay{2007}{}{}.
\newblock
\BBOQ{}\APACrefatitle{{elrm: Software implementing exact-like inference for
  logistic regression models}}{{elrm: Software implementing exact-like
  inference for logistic regression models}}.\BBCQ{}
\newblock
\APACjournalVolNumPages{J. Statist. Software}{21}{}{1--18}.
\PrintBackRefs{\CurrentBib}

\end{thebibliography}

\end{document}